\documentclass[11pt]{article}
\usepackage{listings}
\usepackage{pdflscape}
\usepackage{amsfonts}
\usepackage{epsfig}
\usepackage{lscape}
\usepackage{longtable}
\usepackage{amsmath,amssymb,amsthm}
\usepackage{graphicx}
\usepackage{subfigure}
\usepackage{color}
\usepackage{placeins}
\usepackage{url}
\usepackage{cases}
\usepackage{longtable}
\usepackage{supertabular}
\usepackage{multirow}

\usepackage{amsmath}
\allowdisplaybreaks[4]
\usepackage{cite}
\usepackage{algorithmic}
\usepackage{algorithm,float}
\usepackage{booktabs}
\usepackage{enumitem}
\usepackage{bbm}



\newtheorem{theorem}{Theorem}

\newtheorem{remark}{Remark}
\newtheorem{definition}{Definition}
\newtheorem{proposition}{Proposition}

\usepackage{caption}
\def\R{\mathbb{R}}

\def\N{\mathbb{N}}
\def\E{\mathbb{E}}
\def\P{\mathbb{P}}
\def\bB{\mathbb{B}}

\def\o{\omega}

\def\s{\sigma}

\def\c{\complement}
\def\O{\Omega}

\def\L{\Lambda}
\def\la{\lambda}
\def\t{\theta}

\def\del{\delta}

\def\fM{\mathfrak{M}}
\def\hl{\hat{\lambda}}
\def\hk{\hat{k}}
\def\htt{\hat{\theta}}
\def\cM{\mathcal{M}}
\def\B{\mathbb{B}}

\renewcommand{\l}[0]{\left }
\renewcommand{\r}[0]{\right}

\makeatletter
\renewcommand*{\@cite@ofmt}{\hbox}
\makeatother

\begin{document}

\title{\bf Signal Analysis via the Stochastic Geometry of Spectrogram Level Sets \footnotemark[1]}
\author{Subhroshekhar~Ghosh\footnotemark[2], \quad Meixia~Lin\footnotemark[3], \quad Dongfang~Sun\footnotemark[4]}
\date{\today}
\maketitle

\renewcommand{\thefootnote}{\fnsymbol{footnote}}
\footnotetext[1]{Digital Object Identifier 10.1109/TSP.2022.3153596. S.~Ghosh was partially supported by the MOE grants  R-146-000-250-133, R-146-000-312-114 and MOE-T2EP20121-0013.}
\footnotetext[2]{Department of Mathematics, National University of Singapore, Singapore (email: {\tt subhrowork@gmail.com}; webpage: {\tt https://subhro-ghosh.github.io/}).}
\footnotetext[3]{(Corresponding author) Institute of Operations Research and Analytics, National University of Singapore, Singapore (email: {\tt lin{\_}meixia@u.nus.edu}; webpage: {\tt https://linmeixia.github.io/}).}
\footnotetext[4]{Department of Mathematics, National University of Singapore, Singapore (email: {\tt sun{\_}dongfang@u.nus.edu}).}
\renewcommand{\thefootnote}{\arabic{footnote}}

\begin{abstract}
   Spectrograms are fundamental tools in time-frequency analysis, being the squared magnitude of the so-called short time Fourier transform (STFT). Signal analysis via spectrograms has traditionally explored their  \textit{peaks}, i.e. their maxima. This is complemented by a recent interest in their zeros or minima, following seminal work by Flandrin and others, which exploits connections with Gaussian analytic functions (GAFs). However, the zero sets (or extrema) of GAFs have a complicated stochastic structure, complicating any direct theoretical analysis. Standard techniques largely rely on statistical observables from the analysis of spatial data, whose distributional properties for spectrograms  are mostly understood only at an empirical level. In this work, we investigate spectrogram analysis via an examination of the stochastic geometric  properties of their \textit{level sets}. We obtain rigorous  theorems  demonstrating the efficacy of a spectrogram level sets based approach to the detection and estimation of signals, framed in a concrete inferential set-up. Exploiting these ideas as theoretical underpinnings, we propose a level sets based algorithm for signal analysis that is intrinsic to given spectrogram data, and substantiate its effectiveness via extensive empirical studies. Our results also have theoretical implications for spectrogram zero based approaches to signal analysis. To our knowledge, these results are arguably among the first to provide a rigorous statistical understanding of signal detection and reconstruction in this  set up, complemented with provable guarantees on detection thresholds and rates of convergence.
\end{abstract}

\medskip
\noindent
{\bf Keywords:} Gaussian analytic function, level sets, stochastic geometry, Gabor spectrogram, statistical inference, signal analysis

\section{Introduction}
\label{sec:intro}
Short Time Fourier Transforms (abbrv. STFTs) are foundational objects in time-frequency analysis \cite{Gr01,Fla98,Co12}. By introducing a window function in the traditional Fourier transform, STFTs lead to a two-dimensional representation of a signal on the time-frequency plane. 

In a bit more detail, if $f \in L^2(\R)$ is a signal and $\phi \in L^2(\R)$ is a (real-valued) window function, then the STFT of $f$ with respect to the window function $\phi$, denoted by $V_\phi f$, is a function on $\R^2$ defined as 
\begin{equation*} 
V_\phi f (\tau,\o)=\int_R f(t)\phi(t-\tau) e^{-2i\pi tw} \mathrm{d}t.
\end{equation*}
Informally, the variable $\tau$ is the time variable, whereas $\o$ is the frequency variable in the time-frequency plane. Popular choices for the window function include the Gaussian density function $\phi(t)=\frac{1}{\sqrt{2\pi}\s}e^{-t^2/2\s^2}$. A key effect of such window functions is to localise the signal $f$ in time (e.g., at the scale $\s$ for the Gaussian window above). 

A fundamental quantity that is a functional of the STFT of a signal is the \textit{spectrogram}, which is going to be germane to the investigations carried out in this paper. The spectrogram of a signal is the square of the modulus of the STFT of the signal. To be more precise,  if $f \in L^2(\R)$ is a signal and $\phi \in L^2(\R)$ is a (real-valued) window function, then the spectrogram of $f$ with respect to the window function $\phi$, denoted by $S_\phi f$, is a function on $\R^2$ defined as 
\begin{equation} 
S_\phi f (\tau,\o)=\l|\int_R   f(t)\overline{\phi(t-\tau)} e^{-2i\pi tw} \mathrm{d}t\r|^2 = |V_\phi f  (\tau,\o)|^2.\label{eq: def_spectrogram}
\end{equation}
The specific setting of STFT with a Gaussian window function is referred to as the Gabor transform \cite{Gr01}. Informally speaking, the spectrogram of a signal associating with each point in the time-frequency plane is a localised measure of the \textit{energy} of the signal at that time and that frequency \cite{BarHar21}.

For a more detailed technical discussion on STFTs and their implications for \textit{white noise}, we refer the reader to Section \ref{sec:STFT-WN}. A comprehensive exposition with complete theory and applications is available in \cite{Gr01}.

Spectrogram analysis has been demonstrated to be highly effective in many applications. A particularly significant domain of application is in the field of acoustics. For instance, AM-FM-type signals with a small number of components admit spectrogram representations that are \textit{sparse}, in the sense that locations with high spectral energy are few and far between on the time-frequency plane, and allow for a cogent explanation in terms of the structure of the original signal \cite{Fla15}. We refer the interested reader to \cite{Gr01}, and the references therein, for a more exhaustive overview of the various application domains and their specific problems.

Classically, considerable attention has been focused on the \textit{maxima} of the spectrogram. This is related to the understanding that these capture greater energy of the spectrogram, and therefore greater information about the signal. Techniques such as synchrosqueezing, reassignment and ridge extraction have gained prominence in the context of identifying and processing the maxima of the spectrogram. For details on the classical methods, we refer the reader to the excellent monographs \cite{Gr01,Fla98,Co12,Fla18}.

In recent years, a somewhat different strain of investigations has gained prominence. This pertains to examining the \textit{zeros}, as opposed to the maxima of the spectrogram. In other words, instead of the \textit{high peaks}, the attention shifts to the \textit{deep valleys} of the spectrogram. This line of investigations originates from seminal work of Flandrin \cite{Fla15}. In essence, the central idea in this vein of works revolves around the empirical observation of \cite{Fla15} that the Gabor transform (i.e., STFT with a Gaussian window) of white noise is, in fact, an analytic function on the time-frequency plane (in the single complex variable $\tau+i\o$), and as such, has a well-structured set of zeros that form a discrete set without any accumulation points.  Furthermore, Flandrin observed that these zeros (which are also the zeros of the Gabor spectrogram of white noise) exhibit a spatial distribution that is highly uniform (i.e., grid-like) on the time-frequency plane, with  highly regular Voronoi tessellation patterns. On the other hand, the presence of a non-zero signal creates distortions in the highly uniform spatial distribution of points, which can possibly be exploited for separating signal from noise, and estimating the signal if one is present. 

In a related direction, there has been a recent line of work connecting the STFT of white noise to the so-called Gaussian Analytic Functions (abbrv., GAFs). In particular, it was established in \cite{Bar20} that the zeros of the spectrogram of white noise have the same statistical distribution as the zeros of the planar GAF, which is the most common model of GAFs defined in the setting of the complex plane. GAFs and related models of Determinantal Point Processes (abbrv., DPPs) have attracted great interest in probability theory and statistical physics over the past two decades; see \cite{HKPV,NS10} for a comprehensive overview. Such processes are characterised by the so-called \textit{hyperuniformity} of the spatial distribution of points, which is of independent interest in the statistical physics and condensed matter physics communities \cite{GL17a,GL17b,TorSti03}. 

In subsequent work, the relationship between spectrogram zeros of white noise and zeros of GAFs has been extended to other geometries (apart from the planar case, such as spherical and hyperbolic geometries) \cite{BarHar21}. There is an extensive theory of GAFs and their zeros on these spaces (see, e.g., \cite{HKPV, PerVir05}), which have been connected to zeros of spectrograms of white noise with non-Gaussian windows on the time-frequency analysis side. In particular, the zeros of the hyperbolic GAF have been demonstrated to correspond to the so-called \textit{analytic wavelet transform} of Daubechies and Paul \cite{DauPau88}; in this context we also refer the reader to the work \cite{AHKR18} that appeared in parallel to and independent of \cite{BarHar21}. Further connections to Gaussian analytic functions were investigated in \cite{abreu2020local}.

In the context of Gabor spectrogram zeros of white noise, the hyperuniformity of the GAF zero sets provides a cogent explanation for the empirical observation in \cite{Fla15} that these zeros have a highly homogeneous spatial distribution. However, the procedures to exploit this property in order to perform signal detection and inference are largely empirical in nature. The  mathematically rigorous understanding of why and how such procedures are effective, in terms of theoretical guarantees or quantitative rates, is rather limited.

\begin{figure}[h]
	\centering
	\includegraphics[width=0.35\linewidth]{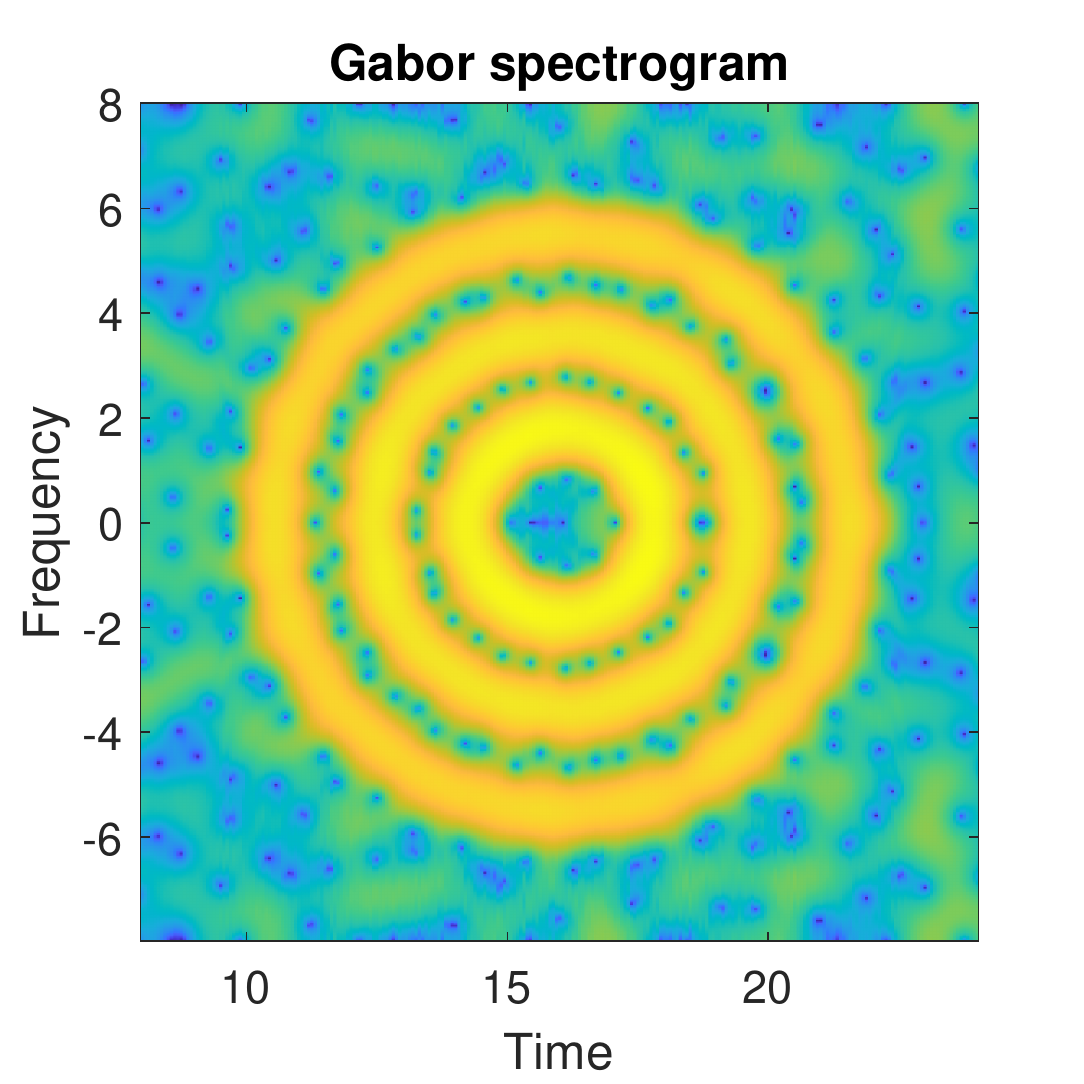}
	\includegraphics[width=0.35\linewidth]{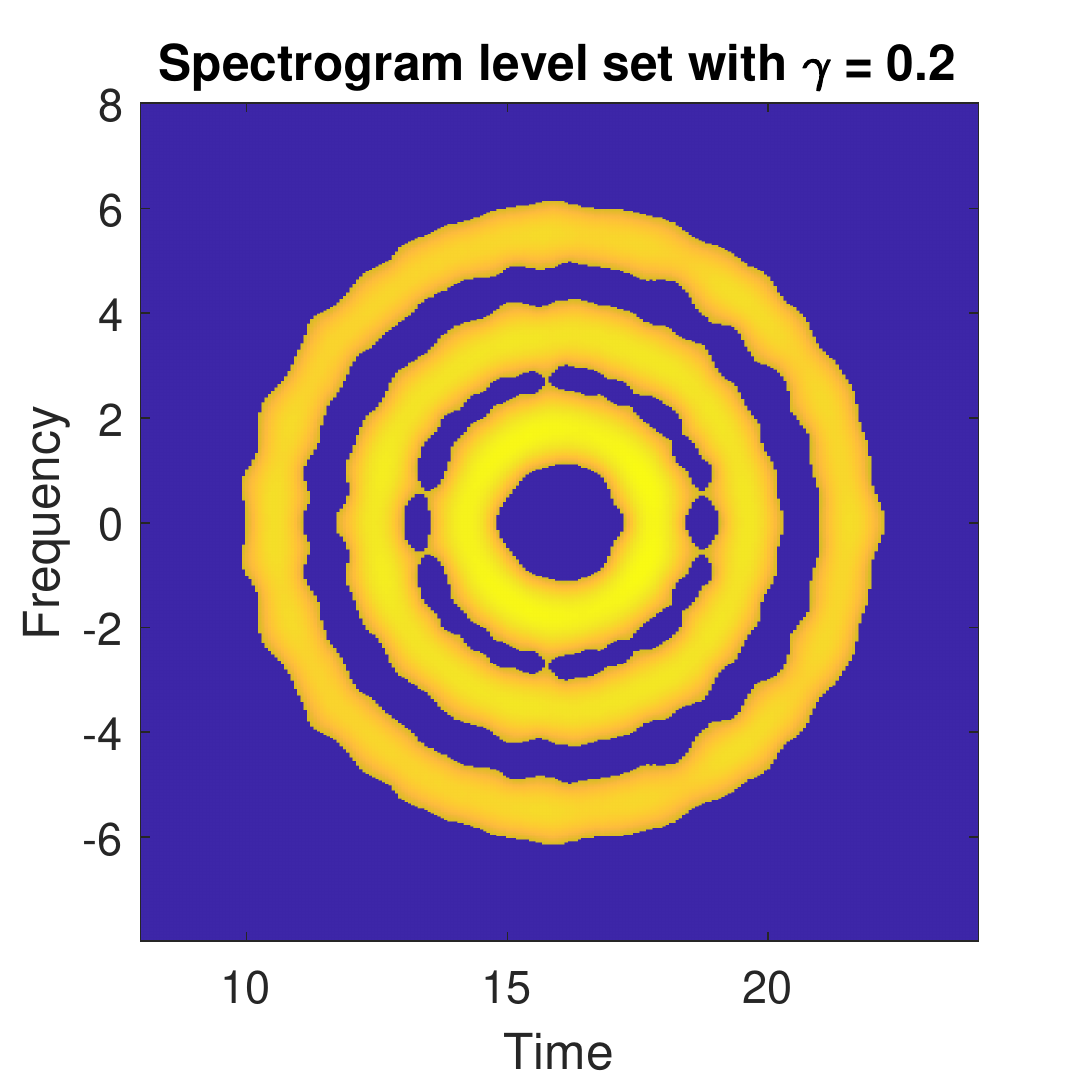}
	\caption{Gabor spectrogram and spectrogram level set of linear combination of fundamental modes corrupted by noise with $k_1=10$, $k_2=40$, $k_3=95$ in \eqref{eq: more_mode}.}\label{fig: Gabor_spectrogram_multiple_10_40_95}
\end{figure}

In this work, we investigate the stochastic geometry of the \textit{level sets} of the Gabor spectrogram of white noise. A \textit{random field} $X$ on a set $E$, roughly speaking, is a collection of (usually real or complex valued) random variables $\{X_v\}_{v \in \E}$ indexed by the elements of $E$ that have a joint statistical distribution. Generally, random fields are studied under regularity conditions with respect to some intrinsic geometry on the set $E$ (such as almost sure continuity or smoothness or H$\ddot{\rm o}$lder regularity, etc.). A typical example of such a background set $E$ would be a Euclidean space $\R^d$ with its canonical Euclidean metric geometry and Lebesgue measure. For any random field $X$ on a set $E$, and a threshold $u \in \R$, the \textit{level set} $\L_X(u)$ is the set 
\begin{equation} \label{eq:def-levset}
\L_X(u):=\{v \in E :  |X_v| \ge u \}.
\end{equation}
In other words, $\L_X(u)$ is the \textit{random} subset of $E$ where the random field $X$ \textit{exceeds} the threshold $u$. To be more precise, $\L_X(u)$ as defined above is sometimes referred to as the \textit{upper level set} for the threshold $u$; whereas the random subset $\L_X(u)^\c \subset E$ is referred to as the lower level set. However, since the understanding of these two sets is more  or less interchangeable for our purposes, we will adhere to the concept of the level set as outlined in \eqref{eq:def-levset}.

In this work, we  undertake a comparative investigation of  the level sets of the Gabor spectrogram of Gaussian white noise and those of a signal corrupted by  Gaussian white noise. In particular, we obtain a mathematically rigorous understanding of certain (random) geometric and analytical properties of such level sets. This analysis is carried out in Section \ref{sec:spec-WN}. Utilising this analysis as a cornerstone, we establish the theoretical foundations of a spectrogram level sets based approach to signal analysis in the time-frequency domain. This is explored in Section \ref{sec:levset-theory}.

To be precise, we frame our result in terms of a natural hypothesis testing problem to decide between pure white noise and the presence of a fundamental signal. We provide an efficient test of hypothesis for this problem, and provide theoretical guarantees for its effectiveness via non-asymptotic bounds on the power of the test. Based on stochastic geometric considerations of spectrogram level sets, we also demonstrate an estimation procedure for a fundamental signal if it is present, and provide  error bounds for the accuracy of such estimation in terms of the quantity of data available. 

Our analysis of level sets also has bearing on the spectrogram zero based approach to filtering (on the lines of \cite{Fla15}), by rigorously demonstrating that the presence of a non-trivial signal creates zero-free regions for the spectrogram on the time-frequency plane, whose geometry is related to the structure of the signal in a systematic manner. The zero sets (or the maxima or minima) of GAFs have a complicated stochastic structure, which makes a direct theoretical analysis of usual spectrogram analysis techniques via GAFs a difficult proposition. 

For inferential purposes, these techniques largely rely on  statistical observables from the analysis of spatial data, whose distributional properties for spectrogram extrema  are mostly understood empirically. Such statistics include, e.g., the celebrated Ripley's K function (cf. \cite{Bar20}). Since the (upper) level sets of the Gabor spectrogram of signals cannot contain zeros of the spectrogram, our results provide theoretical validation for spectrogram zero based techniques, by demonstrating the appearance of canonical zero-free regions on the time-frequency plane when a signal is present along with noise.

Motivated by our theoretical analysis, we propose an algorithm for signal analysis that is intrinsic to the spectrogram data (and, in particular, does not require  any additional inputs regarding  signal strength and such). This procedure is able to effectively perform detection and estimation for linear combinations of fundamental signals. The signal estimation turns out to be highly accurate as long as the fundamental signals being combined are reasonably well separated. We demonstrate the effectiveness of our algorithm via empirical evidence in the form of numerical experiments. The algorithm and its various attributes are analysed in Section \ref{sec:Algo}.

At this point, a few words would be in order regarding the comparison and contrast of our approach vis-a-vis  thresholding-based techniques, which are among the most common denoising methods in signal processing and statistics, see e.g.  \cite{mallat1999wavelet,donoho1995noising,donoho1995wavelet}. It may be noted in this vein that the spectrogram level sets we consider are spatial objects that live on the time-frequency plane while the standard thresholding-based  procedures generally truncate the small coefficients in some basis expansion of an estimator of the signal. Furthermore, the two approaches are conceptually different from an algorithmic point of view. Our approach considers the level set as an input and provides an estimate of the signal as an output (if it detects that one exists). In contrast,  generic thresholding-based methods first find a rough estimate of the signal and then truncate  small coefficients to zero. That is to say, such techniques use thresholding as a step in their algorithm. Last but not least, the coefficients in the standard thresholding methods are finite in number, so  the cost of an inadvertent error in thresholding can be substantial. On the other hand, spectrogram level sets being continuous objects, we believe that our approach  would be more robust to procedural noise. Our approach sheds light on the limiting performance of standard thresholding-based techniques when the discretization parameters tend to zero, which is a common and useful idealisation of spectrogram-based methods in practice.

To the best of our knowledge, our results are arguably among the first to provide a rigorous statistical understanding of signal detection and reconstruction in this  set up, complemented with provable guarantees on detection thresholds and rates of convergence. We therefore focus our attention on the most fundamental setting for a signal, namely Hermite functions, which form an orthonormal basis for $L^2(\R)$ and is central to Gabor analysis; our results appear to be novel even in such a set up. Consideration of more refined signal classes, with characteristic analytical or geometric properties, suggests a natural avenue for future research. 

\section{STFT and white noise}
\label{sec:STFT-WN}

\subsection{The short-time Fourier transform}
\label{sec:STFT}
Given a signal $f$ and a window function $\phi$, the short-time Fourier transform $V_{\phi}f(\cdot,\cdot)$ is defined as the inner product between $f$ and shifted $\phi$ (in the sense of time and frequency). The detailed definition is shown as follows, which could be found in \cite[Definition 3.1.1]{Gr01}.
\begin{definition}\label{def: stft}
	Fix a window function $\phi\in L^2(\mathbb{R})$. The short-time Fourier transform (STFT) of a function $f\in L^2(\mathbb{R})$ with respect to $\phi$ is defined as
	\begin{align}
	V_\phi f(u,v) := \langle f,M_v T_u \phi \rangle,\quad u,v\in \mathbb{R},\label{eq: def_stft}
	\end{align}
	where $M_v f =e^{2i\pi v \cdot} f(\cdot)$ and $T_u f = f(\cdot - u)$. Here $\langle \cdot,\cdot\rangle $ denotes the inner product in $L^2(\mathbb{R})$ with respect to the Lebesgue measure, which is defined as
	\begin{align*}
	\langle f,g\rangle = \int f(t)\overline{g(t)}\mathrm{d}t,\quad f,g\in L^2(\mathbb{R}).
	\end{align*}
\end{definition}

It needs to be mentioned that the above definition is not unique, as there is some flexibility in the choice of the phase term in the integral. For example, the symmetric form $\int  f(t+\tau/2)\overline{\phi(t-\tau/2)} e^{-2i\pi tw} \mathrm{d}t$, which is often called the cross-ambiguity function, plays an important role in radar and in optics \cite{cook2012radar,woodward2014probability}. Introducing a pure-phase term in \eqref{eq: def_stft} has no impact on the corresponding spectrogram but it can help in getting symmetric forms of the STFT, see Proposition \ref{prop: Bargmann}.

\begin{remark}
	Note that the definition of the STFT in \eqref{eq: def_stft} could be extended to more general functions, such as tempered distributions. The detailed discussion is given in Section \ref{sec:WN}. A typical example of tempered distributions is Gaussian white noise.
\end{remark}

It is of special interest to study the STFT with respect to a Gaussian window function $g(t)=2^{1/4}e^{-\pi t^2}$ such that $\|g\|_{L^2(\mathbb{R})}=1$. This special setting of the STFT is referred to as the Gabor transform. The following proposition provides a way to compute the Gabor transform using the Bargmann transform, which is taken from \cite[Proposition 3.4.1]{Gr01}.
\begin{proposition}\label{prop: Bargmann}
	For $f\in L^2(\mathbb{R})$, it holds that
	\begin{align*}
	V_gf(u,v) = \exp(-\pi i u v-\frac{\pi}{2}|z|^2)Bf(\bar{z}),\quad u,v\in \mathbb{R},
	\end{align*}
	where the complex number $z$ is taken as $z=u+iv$, and the Bargmann transform \cite{bargmann1961hilbert} of a function $f$ on $\mathbb{R}$ is the function $Bf$ on $\mathbb{C}$ defined by
	\begin{align*}
	Bf(z)=2^{1/4} \int_{\mathbb{R}} f(t) \exp(2\pi tz-\pi t^2 -\frac{\pi}{2}z^2)\mathrm{d}t,\quad z\in \mathbb{C}.
	\end{align*}
\end{proposition}

Some functions are known to have a simple closed-form Bargmann transform, such as the Hermite functions. Let $\{H_k\}$ be Hermite polynomials \cite{Gr01,holden2010stochastic} defined as
\begin{align*}
H_k(x) = \frac{(-1)^k}{2^{2k-1/4}\pi^k k!}e^{\frac{3}{2}\pi x^2}\frac{\mathrm{d}^k}{\mathrm{d}x^k}\left(e^{-2\pi x^2}\right),\quad  k = 0,1,\cdots,
\end{align*}
which are orthonormal in $L^2(\mathbb{R})$ with respect to the window function $g$, i.e.,
\begin{align*}
\int H_k(x) H_l(x) g(x) \mathrm{d}x = \delta_{kl},\quad k,l\geq 0,
\end{align*}
where $\delta_{kl}$ is the Kronecker delta. The Hermite functions are then defined as: for each $k=0,1,\cdots$,
\begin{align}
h_k(x) = \frac{(-1)^k}{2^{2k-1/2}\pi^k k!}e^{\pi x^2}
\frac{\mathrm{d}^k}{\mathrm{d}x^k}\left(e^{-2\pi x^2}\right).\label{eq: def_hk}
\end{align}
It can be proved that $\{h_k\}$ form an orthonormal basis of $L^2(\mathbb{R})$. According to \cite[Theorem 3.4.2]{Gr01}, the Bargmann transform of the Hermite function $h_k$ is in a quite simple form as
\begin{align*}
Bh_k(z)  = \frac{\pi^{k/2}z^k}{\sqrt{k!}},\quad k = 0,1,\cdots.
\end{align*}

As a side note, we consider the form of Hermite functions in \eqref{eq: def_hk} instead of the canonical form is to help us get a clean formula of its Bargmann transform.

Based on the above discussions and Proposition \ref{prop: Bargmann}, we can see that the Gabor transform of the Hermite function $h_k$ could be computed via the following proposition.
\begin{proposition}\label{prop: stft_hk}
	For any $k\in \mathbb{N}$, it holds that
	\begin{align*}
	V_g h_k (u,v) = \exp(-\pi i u v-\frac{\pi}{2}|z|^2)\frac{\pi^{k/2}\overline{z}^k}{\sqrt{k!}},\quad u,v\in\mathbb{R},
	\end{align*}
	where the complex number $z$ is taken as $z=u+iv$.
\end{proposition}

\subsection{Gaussian white noise}
\label{sec:WN}
In order to define white noise in signal processing, we need the definitions of Schwartz space and tempered distributions \cite{rudin1973,holden2010stochastic}.
\begin{definition}
	Schwartz space ${\cal S}(\mathbb{R})$ is the function space consisting of rapidly decreasing smooth functions from $\mathbb{R}$ to $\mathbb{C}$, i.e.,
	\begin{align*}
	{\cal S}(\mathbb{R})=\left\{ \phi\in C^{\infty}(\mathbb{R},\mathbb{C})\Big| \sup_{x\in \mathbb{R}}
	|x^{\alpha}\phi^{(\beta)}(x)|<\infty,\ \forall \alpha,\beta\in \mathbb{N}\right\}.
	\end{align*}
\end{definition}

\begin{definition}
	The space of tempered distributions on $\mathbb{R}$, denoted as ${\cal S}'(\mathbb{R})$, is the continuous dual of ${\cal S}(\mathbb{R})$, which consists of all linear and continuous functions from ${\cal S}(\mathbb{R})$ to $\mathbb{C}$.
\end{definition}
From the definition of Schwartz space, we can see that ${\cal S}(\mathbb{R})$ is a dense subspace of $L^2(\mathbb{R})$. Furthermore, we have that ${\cal S}(\mathbb{R})\subseteq L^2(\mathbb{R})\subseteq {\cal S}'(\mathbb{R})$. For any $\psi\in S'(\mathbb{R})$, $\phi\in S(\mathbb{R})$, we define the action $\langle \psi,\phi\rangle:=\psi(\phi)$. Based on this notation, the definition of STFT in \eqref{eq: def_stft} could be naturally extended. That is, the STFT of $f\in {\cal S}'(\mathbb{R})$ with respect to a window function $\phi\in {\cal S}(\mathbb{R})$ is 
\begin{align*}
V_\phi f(u,v) := \langle f,M_v T_u \phi \rangle,\quad u,v\in \mathbb{R}.
\end{align*}

Now we could define the white noise in a classical approach. We use the space of tempered distributions ${\cal S}'(\mathbb{R})$ as the basic probability space, and its Borel subsets, denoted as ${\cal B}({\cal S}'(\mathbb{R}))$, as the event space. According to the Bochner-Minlos theorem \cite[Theorem 2.1.1]{holden2010stochastic}, there exists a unique probability $\mu_1$ on ${\cal B}({\cal S}'(\mathbb{R}))$ with the following property:
\begin{align*}
\mathbb{E}\left[e^{i\langle \cdot,\phi\rangle}\right]:=\int_{S'(\mathbb{R})}
e^{i\langle \xi,\phi\rangle}\mathrm{d}\mu_1(\xi)=e^{-\frac{1}{2}\|\phi\|^2},\quad  \phi \in {\cal S}(\mathbb{R}),
\end{align*}
where $\|\phi\|^2=\|\phi\|^2_{L^2(\mathbb{R})}$, $\langle \xi,\phi\rangle=\xi(\phi)$ is the action of $\xi\in S'(\mathbb{R})$ on $\phi\in S(\mathbb{R})$ and $\mathbb{E}=\mathbb{E}_{\mu_1}$ denotes the expectation with respect to $\mu_1$. The triplet $(S'(\mathbb{R}),{\cal B}(S'(\mathbb{R})),\mu_1)$ is called the while noise probability space, and $\mu_1$ is called the white noise measure. The measure $\mu_1$ is also called the normalized Gaussian measure on ${\cal S}'(\mathbb{R})$, since according to
\cite[Lemma 2.1.2]{holden2010stochastic}, for any random variable $\xi$ with distribution $\mu_1$ and $\phi_1,\cdots,\phi_n \in {\cal S}(\mathbb{R})$ that are orthonormal in $L^2(\mathbb{R})$, the vector $(\langle \xi,\phi_1\rangle,\cdots,\langle \xi,\phi_n\rangle)$ is a real standard Gaussian random vector. 

Let $\xi$ be a random variable with distribution $\mu_1$. The following proposition identifies each value of the random function ${\cal F}[\xi](z):=V_g \xi(u,v)$, i.e., the Gabor transform of $\xi$, as a limit in $L^2(\mu_1)$, which comes from \cite[Proposition 2 and Proposition 3]{Bar20}.

\begin{proposition}\label{prop: entire}
	Let $u,v\in \mathbb{R}$ and write $z=u+iv\in \mathbb{C}$. Then 
	\begin{align}
	{\cal F}[\xi](z)= \sqrt{\pi} \exp(i\pi u v-\frac{\pi}{2}|z|^2) \sum_{k=0}^{\infty} \langle \xi,h_k\rangle \frac{\pi^{k/2} z^k}{\sqrt{k!}},\label{eq: def_Fxi}
	\end{align}
	where $\{h_k\}$ are the orthonormal Hermite functions defined in \eqref{eq: def_hk}, and convergence is in $L^2(\mu_1)$. In addition, the random series 
	\begin{align*}
	\sum_{k=0}^{\infty} \langle \xi,h_k\rangle \frac{\pi^{k/2} z^k}{\sqrt{k!}}
	\end{align*}
	$\mu_1$-almost surely defines an entire function.
\end{proposition}

It should be noted that in the remaining part, we use the notation $V_g \xi(\cdot,\cdot)$ when we treat the function in \eqref{eq: def_Fxi} as a function on $\mathbb{R}^2$, and use the notation ${\cal F}[\xi](\cdot)$ when we treat the function in \eqref{eq: def_Fxi} as a function on $\mathbb{C}$.

\section{The spectrogram of white noise}
\label{sec:spec-WN}
The squared modulus of the STFT is called the spectrogram, which is commonly interpreted as a measure of the signal around time and frequency. In order to identify fundamental modes, we need the boundedness of the Gabor spectrogram (i.e., the spectrogram of the STFT with a Gaussian window) of white noise, within a bounded set, which enables us to obtain a mathematically rigorous understanding of certain (random) geometric and analytical properties of spectrogram level sets. 

\subsection{Gaussian random fields}
Note that $\{F[\xi](z)\}_{z\in \mathbb{C}}$ is a centered Gaussian random field indexed by $\mathbb{C}$, which is determined by its covariance function. The covariance function of $\{F[\xi](z)\}_{z\in \mathbb{C}}$ could be computed as in the following proposition, whose proof is in Section \ref{sec:covariance}.
\begin{proposition}\label{prop: covariance}
	For any $z,w\in \mathbb{C}$, we write them in the form as $z=u_1+iv_1$, $w=u_2+iv_2$. The covariance function of $\{F[\xi](z)\}_{z\in \mathbb{C}}$ is
	\begin{align*}
	K(z,w) &=\mathbb{E}\left[\left( {\cal F}[\xi](z)-\mathbb{E}\left[{\cal F}[\xi](z)\right]\right) \overline{\left( {\cal F}[\xi](w)-\mathbb{E}\left[{\cal F}[\xi](w)\right]\right) }\right]\\
	&= \pi \exp(i\pi u_1v_1-i\pi u_2v_2-\frac{\pi}{2}|z|^2-\frac{\pi}{2}|w|^2+\pi z\overline{w}).
	\end{align*}
	Moreover, the variance function of $\{F[\xi](z)\}_{z\in \mathbb{C}}$ is
	\begin{align*}
	\sigma^2(z):= \mathbb{E}\left[|{\cal F}[\xi](z)-\mathbb{E}\left[{\cal F}[\xi](z)\right]|^2\right]=\pi.
	\end{align*}
\end{proposition}

Before describing our main result, we need a metric $d$ on $\mathbb{C}$, known as the canonical metric for ${\cal F}[\xi](\cdot)$, by setting
\begin{align*}
d(z,w)=\left\{ \mathbb{E}\left[|F[\xi](z)-F[\xi](w)|^2\right]\right\}^{1/2},\quad z,w\in \mathbb{C}.
\end{align*}
The following proposition provides the computation of $d(\cdot,\cdot)$ in an explicit formula, and the proof could be referred to Section \ref{sec: proof_d}.
\begin{proposition}\label{prop: d}
	For any $z=u_1+i v_1$, $w=u_2+i v_2\in \mathbb{C}$, it holds that
	\begin{align*}
	d^2(z,w)=2\pi \left[1-\cos  \left(\pi(u_1+u_2)(v_1-v_2) \right)\exp(-\frac{\pi}{2}|z-w|^2)\right].
	\end{align*}
\end{proposition}

\subsection{The Gabor spectrogram of white noise}
\label{sec:max-WN}
In this subsection, we focus on the boundedness of the Gabor spectrogram of white noise $\{F[\xi](z)\}_{z\in \mathbb{B}_L}$, where the parameter set $\mathbb{B}_L$ is the ball in $\mathbb{C}$ with size $L>0$ defined as
\begin{align}\label{eq: def_BL}
\mathbb{B}_L := \{u+iv\in \mathbb{C}: \max\{|u|,|v|\}\leq L, u,v\in \mathbb{R} \}.
\end{align}
It can be seen later that the boundedness depends on how large the parameter set is when the size is measured by a metric that comes from the process itself. We measure the size of the parameter set $\mathbb{B}_L$ via the metric entropy, whose definition is given as follows.

\begin{definition}
	Suppose $\{f(t)\}_{t\in T}$ is a centered real valued Gaussian random field with a parameter set $T\subseteq \mathbb{R}$, and $d_f$ is the associated canonical metric defined as $d_f(t,w)=\left\{ \mathbb{E}\left[|f(t)-f(w)|^2\right]\right\}^{1/2}$. Assume that $T$ is $d_f$-compact. Denote the $d_f$ ball centered on $t\in T$ and of radius $\varepsilon$ as
	\begin{align*}
	B_{d_f}(t,\varepsilon):= \{w\in T\mid d_f(t,w)\leq \varepsilon\}.
	\end{align*}
	The metric entropy function for $\{f(t)\}_{t\in T}$ is defined as $N(T,d_f,\varepsilon)$, which is the smallest number of such balls that cover $T$. 
\end{definition}

Note that in the above definition, the real valued Gaussian random field is considered, while our Gaussian random field $\{F[\xi](z)\}_{z\in\mathbb{C}}$ is complex valued. Thus the boundedness of $\{F[\xi](z)\}_{z\in\mathbb{B}_L}$ should be analysed through its real part and imaginary part. The following theorem states that the Gabor spectrogram of white noise is of order $\log{L}$ over the bounded set $\mathbb{B}_L$ with high probability.

\begin{theorem}\label{thm: boundedness_white_noise}
	For $L\geq\pi$, we have that for any $\tau>0$,
	\begin{align*}
	\mathbb{P}\left[ \sup_{z\in \mathbb{B}_L}|{\cal F}[\xi](z)|\leq \sqrt{2}(14K + \tau)\sqrt{\log L}\right]
	\geq 1-4\exp\left(-\frac{\tau^2}{2\pi} \cdot \log L\right),
	\end{align*}
	where $K>0$ is a constant.
\end{theorem}

We defer the proof of Theorem \ref{thm: boundedness_white_noise} to Section \ref{sec: boundedness_white_noise}. Theorem \ref{thm: boundedness_white_noise} plays an important role in investigating the random geometric and analytical property of the spectrogram level sets of fundamental modes corrupted with white noise.

\subsection{Spectrogram level sets of a noisy fundamental mode}
Before investigating the property of the spectrogram level sets of a noisy fundamental mode, we need the following proposition, which shows the maximum value of the Gabor spectrogram of the Hermite function $h_k$ defined in \eqref{eq: def_hk}. The proof could be found in Section \ref{sec:max_spec_signal}.

\begin{proposition} \label{prop:max_spec_signal}
	Given $k\in \mathbb{N}$, let $h_k$ be the Hermite function defined in \eqref{eq: def_hk}. Then it holds that
	\begin{align*}
	\max_{u,v\in \mathbb{R}}|V_g h_k (u,v)| = \prod_{t=1}^k \sqrt{\frac{k}{et}},
	\end{align*}
	where the maximum value is obtained when $\sqrt{u^2+v^2}=\sqrt{k/\pi}$.
	
	Furthermore, for any $u,v\in \mathbb{R}$ such that $\sqrt{u^2+v^2}-\sqrt{k/\pi}=r\sqrt{k/\pi}$ with $r\in \mathbb{R}$, we have
	\begin{align}
	\frac{|V_g h_k (u,v)|}{\max_{u,v\in \mathbb{R}}|V_g h_k (u,v)|}=\frac{(1+r)^k}{e^{k(r+r^2/2)}}.\label{eq: max_ratio}
	\end{align}
\end{proposition}

Denote the spectrogram level set of a signal $y$, restricted to the ball $\bB_L \subset \mathbb{C}$ defined in \eqref{eq: def_BL}, with threshold $\gamma\in \mathbb{R}$ as:
\begin{align}
\Lambda(\gamma):=\{(u,v)\in \mathbb{R}^2:u+iv\in \mathbb{B}_L,\ |V_g y(u,v)|\geq \gamma\}.\label{eq: level_threshold}
\end{align}
We are now ready to investigate the  structure of the level sets of the Gabor spectrogram of a fundamental mode corrupted by additive white noise. The following theorem gives us a mathematically rigorous understanding of geometric and analytical properties of such level sets, with its proof given in Section \ref{sec:stoch-geom-signal}.

\begin{theorem} \label{thm:stoch-geom-signal}
	Assume the signal $y$ is generated as
	\begin{align*}
	y=\la h_k+ \xi,
	\end{align*}
	where $h_k$ is an Hermite function defined in \eqref{eq: def_hk}, white noise $\xi$ is a random variable with distribution $\mu_1$, and $ \la \in \R$, is a parameter. Then it can be seen that $y\in {\cal S}(\mathbb{R})$. Consider the Gabor  spectrogram of $y$ restricted to $\mathbb{B}_L$ as in \eqref{eq: def_BL}. Let 
	\begin{equation*} 
	|\la|\geq \frac{5\sqrt{2}(14K+ \tau)\sqrt{\log L}}{\prod_{t=1}^k \sqrt{k/(et)}},
	\end{equation*}
	where $K>0$ is the constant in Theorem \ref{thm: boundedness_white_noise} and $\tau>0$ is a parameter. Then, for $L\geq \max\{\sqrt{k/\pi},\pi\}$, we have
	\begin{align*} 
	\emptyset \neq \Lambda\left( 3\sqrt{2}(14K + \tau)\sqrt{\log L}\right) \subseteq \left\{(u,v)\in \mathbb{R}^{2}: \frac{|V_g h_k (u,v)|}{\prod_{t=1}^k \sqrt{k/(et)}}> \alpha\right\},
	\end{align*}
	with probability $ \ge 1-4\exp\left(-\frac{\tau^2}{2\pi} \cdot \log L\right)$, where
	\begin{equation*}
	\alpha:=\frac{\sqrt{2}(14K+\tau)\sqrt{\log L}}{|\la|\prod_{t=1}^k \sqrt{k/(et)}}\in (0,\frac{1}{5}].
	\end{equation*}
\end{theorem}

Up to now, we have proved that a well-chosen level set of the Gabor spectrogram of a fundamental mode corrupted by Gaussian white noise is nonempty with high probability, which makes it possible to conduct signal detection via the level sets of the spectrogram of data.

\section{Signal analysis via spectrogram level sets} 
\label{sec:levset-theory}

In this section, we examine the implications of the geometric and analytic properties of spectrograms, as studied in Section \ref{sec:spec-WN}, to the problem of signal analysis. We will focus on the setting of a single fundamental signal corrupted by Gaussian white noise, which offers the requisite structural clarity for obtaining a complete theoretical understanding of the signal analysis problem, at the same time capturing foundational  features of the spectrogram level sets perspective on this question. 

\subsection{The generative and observational models} \label{sec:gen_obs_model}
Herein, we clarify the generative and observation models under which we will work for the rest of Section \ref{sec:levset-theory}.

\begin{itemize}
	\item (Generative model): The observation $y$ is generated as
	\begin{align*}
	y= \la h_k+ \xi,
	\end{align*}
	where the signal $h_k$ is an Hermite function defined in \eqref{eq: def_hk}, white noise $\xi$ is a random variable with distribution $\mu_1$, and $\la$ is the signal strength. We further assume that $1\le k \le k_0$, where $k_0 \in \mathbb{N}$ is given.
	\item (Observational model): The level sets of the spectrogram at any prescribed level $\t$ may be observed, restricted to the $L_\infty$ ball $\bB_L \subset \mathbb{C}$ of radius $L$ defined in \eqref{eq: def_BL}, i.e., the set $\L(\t)$ defined in \eqref{eq: level_threshold}.
\end{itemize}

\subsection{Signal Detection} \label{sec:signal detection}
We will structure our result for signal detection in the form of a hypothesis testing problem as discussed below. The signal detection problem consists in investigating the signal strength $|\lambda|$, relative to the noise (of strength $1$), that is necessary for detecting the presence of the signal. 

\subsubsection{A hypothesis testing perspective} \label{sec:hyp_test_perspective}
We frame the signal detection problem in terms of a natural simple versus composite hypothesis testing question. To be precise, we desire to test the following null vs. alternative hypotheses:
\begin{itemize}
	\item [$H_0$:] The observation $y= \xi$, i.e., there is no signal but only pure noise
	\item [] \textit{vs.} 
	\item [$H_1$:] The observation $y= \la h_k+ \xi$ with $\la \ne 0$ and some integer $k \in [0,k_0]$.
\end{itemize}
A \textit{test} $\psi_\t$ is a measurable function of the observed level set $\L(\t)$ that maps $\L(\t)$ to the (two-element) set $\{0,1\}$ (with the understanding that, the value $0$ corresponds to acceptance of $H_0$ whereas the value $1$ pertains to rejecting the null and accepting the alternative $H_1$). Let $\P^0$ denote the distribution of $\L(\t)$ under $H_0$, whereas $\P_m$ denote the distribution of $\L(\t)$ under $H_1$ with $k=m$.  We say that we \textit{detect} the signal at strength $\la$ if, for any $\del>0$, there exists a test $\psi_{\t(\del)}$ such that \[  \P^0[\psi_{\t(\del)}=1] \vee \max_{1 \le k \le k_0}\P_k[\psi_{\t(\del)} = 0] \le \del. \]
It may be noted that the quantities $ \P^0[\psi_{\t(\del)}=1]$ and $\max_{1 \le k \le k_0 }\P_k[\psi_{\t(\del)} = 0]$ pertain to the probabilities of Type I and Type II errors in this model, and we desire to detect the signal such that the probabilities of there two types of errors are less than $\del$.  

\subsubsection{The test $\psi_\t$} \label{sec:test}
We consider the test 
\begin{equation} \label{eq:test}
\psi_\t = \mathbbm{1}\{ \L(\t) \text{ is non-empty}\}.
\end{equation}
For any $l \ge 0$, we define the minimax quantity 
\begin{equation} \label{eq:minimax}
\fM(l) = \min_{1 \le k \le l} \max_{u,v\in \mathbb{R}}|V_g h_k (u,v)| =  \min_{1 \le k \le l} \prod_{t=1}^k\sqrt{k/(et)}.
\end{equation}
We are now ready to state the following theorem with its proof given in Section \ref{sec:hyp-test}, which provides the theoretical guarantees for the proposed test of hypothesis via non-asymptotic bounds on the power of test.

\begin{theorem} \label{thm:hyp-test}
	In the generative and observational models as in Section \ref{sec:gen_obs_model}, we consider the hypothesis testing problem in Section \ref{sec:hyp_test_perspective}. For any given $\del>0$, consider the test $\psi_{\t(\del)}$ as in \eqref{eq:test} obtained by setting 
	\[\t(\del)=3\sqrt{2}\l(14K \sqrt{\log L} + \sqrt{2\pi \cdot \log (4/\del)} \r),\] 
	where $K>0$ is the constant in Theorem \ref{thm: boundedness_white_noise}. Then, for $L\geq \max\{\sqrt{k_0/\pi},\pi\}$, this test $\psi_{\t(\del)}$ performs \textit{signal detection} as defined in Section \ref{sec:hyp_test_perspective} at signal strength 
	\begin{align}
	|\la| \ge 5\sqrt{2} \ \fM(k_0)^{-1} \l(14K \sqrt{\log L} + \sqrt{2\pi \cdot \log (4/\del)}\r).\label{eq: signal_strengh_detection}
	\end{align}
\end{theorem}

In addition to the hypothesis testing procedure stated above, we also investigate an estimation procedure for a fundamental mode if it is present, with comprehensive theoretical understandings.

\subsection{Signal estimation} \label{sec:signal estimation}
In this section, we demonstrate that signal estimation is possible with high probability (as the observation size $L \to \infty$) in the set up of Section \ref{sec:gen_obs_model}. To this end, we define the following statistics: 
\begin{definition} \label{def:statistics}
	We define
	\begin{itemize}
		\item $\htt:=\max\{\t: \psi_\t =1; \t \ge 0\}$, where $\psi_\t$ is as in \eqref{eq:test};
		\item $\hk:=[[\pi \cdot (\min\{|z|: z \in \L(\htt) \})^2]]$, where $[[x]]$ denotes the nearest integer to $x \in \R$;
		\item $\hl:=\htt / \prod_{t=1}^{\hat{k}}\sqrt{\hat{k}/(et)}$.
	\end{itemize}
\end{definition}

We further introduce a few notations. First, we recall \eqref{eq: max_ratio} and define
\begin{equation*} 
C(k,r) := \frac{(1+r)^k}{ e^{k(r+r^2/2)} },\quad k \in \N,\ r>-1.
\end{equation*}
For every $k \in \N_{+}$, we continue with 
\begin{equation} \label{eq:beta}
\beta(k):= \max\l\{C(k,\frac{1}{4k+2}) , C(k,-\frac{1}{4k+2})\r\}.
\end{equation}
Observe that $\beta(k)<1$ for every $k \in \N_{+}$. Finally, for any given $k_0 \in \N_{+}$, we define 
\begin{equation} \label{eq:cknot}
C(k_0) = \frac{2}{5\left(1-\max_{1 \le k \le k_0} \beta(k)\right)}+1.
\end{equation}
We then state our mode estimation theorem as follows, whose proof is given in Section \ref{sec:estimation-k}. Note that this theorem provides an error bound for the accuracy of the mode obtained by our estimation in terms of the quality of data available.

\begin{theorem} \label{thm:estimation-k}
	In the set-up of Section \ref{sec:gen_obs_model} and parameter $\tau>0$, for $L\geq \max\{\sqrt{k_0/\pi},\pi\}$,  and  signal strength 
	\begin{equation*}
	|\la|\ge t_{\mathrm{mode}}(\tau) := 5 C(k_0) \sqrt{2}(14K+ \tau)\fM(k_0)^{-1}\sqrt{\log L}, 
	\end{equation*} 
	where $K>0$ is the constant in Theorem \ref{thm: boundedness_white_noise}, we have \[\inf_{1 \le k \le k_0}\P_{k}[\hk=k] \ge 1 - 4\exp\left(-\frac{\tau^2}{2\pi} \cdot \log L\right).  \]
\end{theorem}

In addition to the error bound for the estimated mode, we also provide an estimation theorem for signal strength, which demonstrates that our estimation procedure could give a highly accurate estimation of signal strength with high probability. The proof is deferred to Section \ref{sec:estimation-lambda}.
\begin{theorem} \label{thm:estimation-lambda}
	In the set-up of Section \ref{sec:gen_obs_model} and parameter $1\ge \del>0$, for $L> \max\{\sqrt{k_0/\pi},\pi,\exp(14K/\delta)^2\}$, and signal strength 
	\[|\la|\ge t_{\mathrm{strength}} := 5 C(k_0) \sqrt{2} \quad \fM(k_0)^{-1} \log L,\]  
	where $K>0$ is the constant in Theorem \ref{thm: boundedness_white_noise}, we have
	\begin{align*}
	\inf_{1 \le k \le k_0}\P_k\l[\l| \frac{\hl}{|\la|} - 1\r| \le \del \r] 
	\ge 1 - 4\exp\left(-\frac{\del^2 \log^2 L}{2\pi} \cdot \l(1-\frac{14K}{\del \sqrt{\log L}}\r)^2 \right). 
	\end{align*}
\end{theorem}

\section{Algorithm for signal learning via spectrogram level sets}
\label{sec:Algo}
In this section, we propose a spectrogram level sets based algorithm for signal detection and estimation. The algorithm is based on the analysis of geometric and analytic properties of the white noise spectrogram in Section \ref{sec:spec-WN}, and the theoretical investigation of their implications for signal analysis done in Section \ref{sec:levset-theory}. 

\begin{algorithm}[!h]
	\caption{\small {\bf : Signal learning via spectrogram level sets}}
	\label{alg: one_signal}
	\begin{algorithmic}
		\STATE {\bfseries Step 1}. Consider the setting as stated in Section \ref{sec:gen_obs_model}. Compute the Gabor spectrogram of a given signal $y$ with respect to the Gaussian window function $g$, which is denoted as $V_g y(u,v)$, $u,v\in \mathbb{R}$;
		
		\STATE {\bfseries Step 2}. Given $L>0$ large enough, compute
		\begin{align*}
		m_{L}:=\max_{u+iv\in \mathbb{B}_L}|V_g y(u,v)|;
		\end{align*}
		
		\STATE {\bfseries Step 3}. Compute the spectrogram level set 
		\begin{align*}
		\Lambda(0.2 m_{L}) =\left\{(u,v)\in \mathbb{R}^2\mid u+iv\in \mathbb{B}_L, |V_g y(u,v)|\geq 0.2m_{L}
		\right\};
		\end{align*}
		
		\STATE {\bfseries Step 4}. Detect whether these exists an annulus in $\Lambda(0.2 m_{L})$ centered at the origin. 
		\begin{itemize}
			\item[4.1] If so, we say that there exists a fundamental mode.
			\item[4.2] Moreover, suppose the average of the radii of the large circle and the smaller one is $\eta$, the unknown index $k$ could be estimated as $\lfloor{\pi \eta^2}\rfloor$,  where $\lfloor{x}\rfloor$ denotes the greatest integer less than or equal to $x \in \R$;
		\end{itemize}
	\end{algorithmic}
\end{algorithm}

A key feature of the algorithm is that its operation is intrinsic to the spectrogram data, and in particular, does not depend on the prior knowledge of additional information (e.g., regarding the  choice of threshold for the level sets, or otherwise). It also utilizes a single level set of the spectrogram, and does not entail the possibility of having to compare multiple spectrograms (which might be necessary, e.g., to determine $\max\{\t : \psi_\t =1\}$).

A few words are in order regarding the choice of the factor ``0.2". Intuitively, we are going to pick a factor that is intrinsic to the spectrogram data, which can not be close to neither 1 nor 0 as we would like to observe an annulus if there exists a fundamental mode. Empirical evidence shows that ``0.2" is a quite good choice, which provides satisfactory performance for most of the cases. 

As for the procedure to detect the annulus and then estimate the average radii, we give a bit more implementation details in the following remark.
\begin{remark}
	In practice, we always work on a set of grid points $\{(u_i,v_j)\}_{i,j=1}^{2N+1}$ where $u_i=(i-1-N)L/N$, $v_j= (j-1-N)L/N $ and $N>0$ is a given integer. We could construct a $0$-$1$ matrix $S\in \mathbb{R}^{(2N+1)\times (2N+1)}$ as $S_{ij}=1$ if $(u_i,v_j)\in \Lambda(0.2 m_{L})$, and $S_{ij}=0$ otherwise.
	
	Choose a set of slopes $\{\kappa_m\}_{m=1}^M$ with $-\infty<\kappa_1\leq\cdots\leq \kappa_M<\infty$. For each $\kappa_m$, the line $v=\kappa_m u$ approximately intersects the grid at $\{(u_i,v_{j_i^m})\}_{i=1}^{2N+1}$ with $j_i^m:=[[k_m u_i N/L]]+1+N$. Define $b^m\in \mathbb{R}^{2N+1}$ as $b_i^m=S_{u_i,v_{j_i^m}}$, which is a $0$-$1$ vector. Let $D^m$ be the set of distances between grid points indexed by the start points and the end points of the subvector of $b_i^m$ with consecutive elements to be $1$. Set $D=\cup_{i=1}^{M}D^m$. We then apply histogram bin counts on $D$ to see whether there exists a count in one bin exceeds $3M/2$. If so, we say that there exists a fundamental mode, and $\eta$ in the algorithm could be calculated as the median of the bin.
\end{remark}

As a side note, the procedure in Algorithm \ref{alg: one_signal} could be extended to learn linear combinations of fundamental modes as long as the signals being combined are reasonably well separated.

The effectiveness of Algorithm  \ref{alg: one_signal} for learning linear combinations of fundamental modes, can be demonstrated via detailed empirical studies, which we undertake in the following section. 

\section{Empirical investigations}
We conduct empirical studies in this section to demonstrate the performance of our algorithm on detection and estimation for linear combinations of fundamental signals.

In order to measure the performance of Algorithm \ref{alg: one_signal} for learning linear combinations of fundamental signals $\sum_{i=1}^m {\lambda_i}h_{k_i}$ corrupted with noise, where $1\leq k_i\leq k_0$, we need to provide a reasonable metric. It is known that the fundamental modes corresponding to $k$ and $k+1$ are quite similar as functions when $k$ becomes large, but this is no longer very applicable if $k$ is very small, like $1$ or $2$. In order to take this into consideration, we define modified accuracy (mACC) of the estimation $\{\hat{k_j}\}_{j=1}^{\hat{m}}$ as
\begin{align*}
\mbox{mACC} =\left\{\begin{aligned}
&0, && \mbox{if $\hat{m}\neq m$} \\
&\max\left\{0,1-\sum_{i=1}^m \left|\frac{\hat{k_i}}{k_i}-1\right|\right\} && \mbox{if $\hat{m}= m$, $\max_{i}|\hat{k_i}-k_i|\leq  1$},\\
&0, && \mbox{if $\hat{m}= m$, $\max_{i}|\hat{k_i}-k_i|> 1$}.
\end{aligned}\right.
\end{align*}
By definition, we can see that $\mbox{mACC}=1$ means the perfect estimation, and $\mbox{mACC}=0$ represents the worst.

\textbf{One fundamental mode.} We first consider the case of one fundamental mode as stated in Section \ref{sec:gen_obs_model}. That is the generative model is given as
\begin{align}
y=\lambda h_k+\sigma \xi,\label{eq: one_mode}
\end{align}
where the signal $h_k$ is an Hermite function defined in \eqref{eq: def_hk}, white noise $\xi$ is a random variable with distribution $\mu_1$, $\lambda$ is the signal strength and $\sigma$ is the noise strength. In each trail, we 
\begin{itemize}
	\item take the observation radius $L=8$;
	\item randomly generate the integer $k$ uniformly from $\{1,2,\cdots,112\}$;
	\item randomly generate the parameter $\lambda$ uniformly in $[1,2]$;
	\item take the parameter $\sigma$ as $\frac{1}{10\sqrt{\log L}}$.
\end{itemize}
Based on the Gabor spectrogram of the data $y$, we define the spectrogram level set with threshold $\gamma m_L$ as
\begin{align*}
\Lambda(\gamma m_L)=\left\{(u,v)\in \mathbb{R}^2\mid u+iv\in \mathbb{B}_L, \ |V_g y(u,v)|\geq \gamma m_L\right\},
\end{align*}
where $m_L:=\max_{u+iv\in \mathbb{B}_L} |V_g y(u,v)|$. The parameter $\gamma$ is typically taken to be $0.2$ as in our algorithm design.

\begin{figure}[H]
	\centering
	\includegraphics[width=0.35\linewidth]{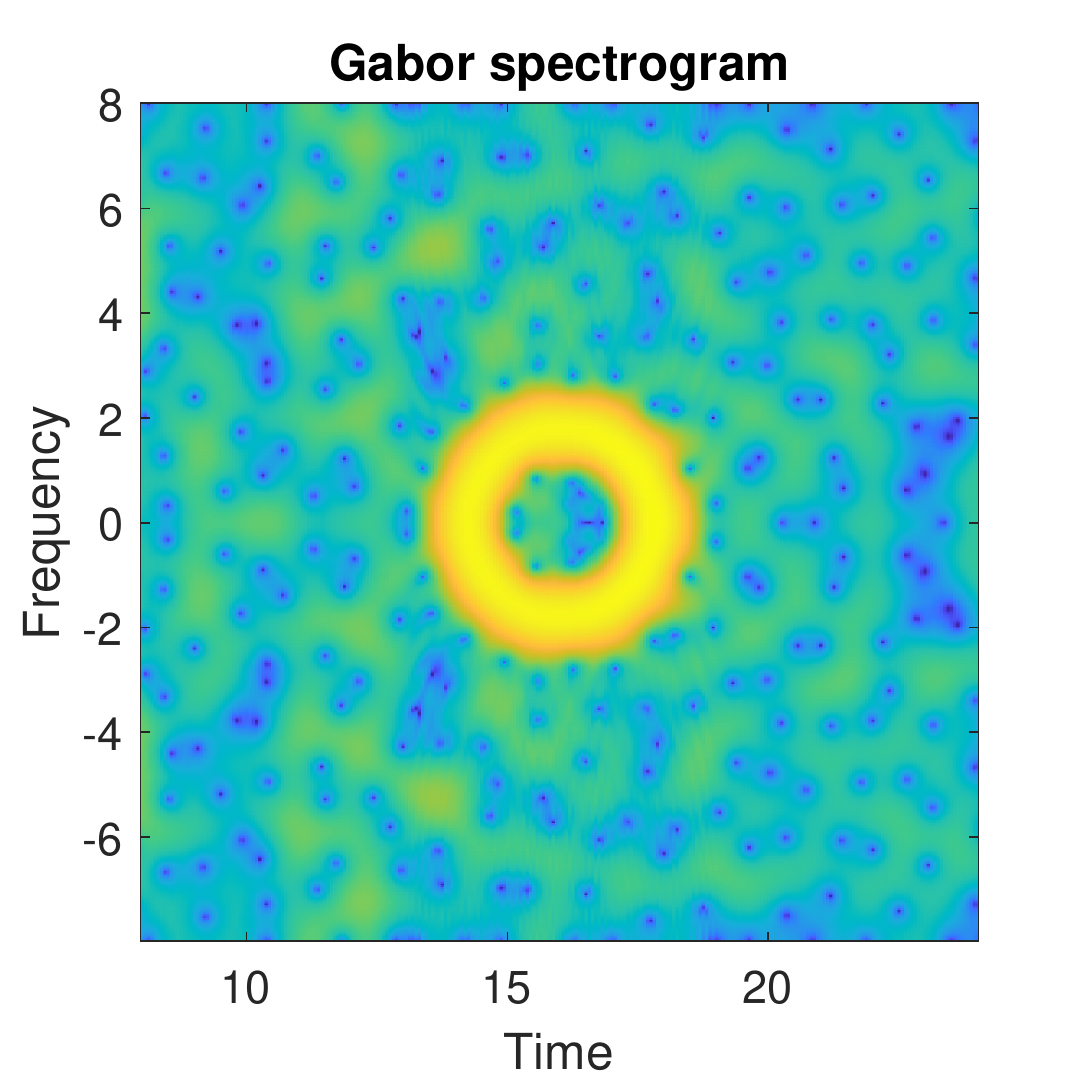}
	\includegraphics[width=0.35\linewidth]{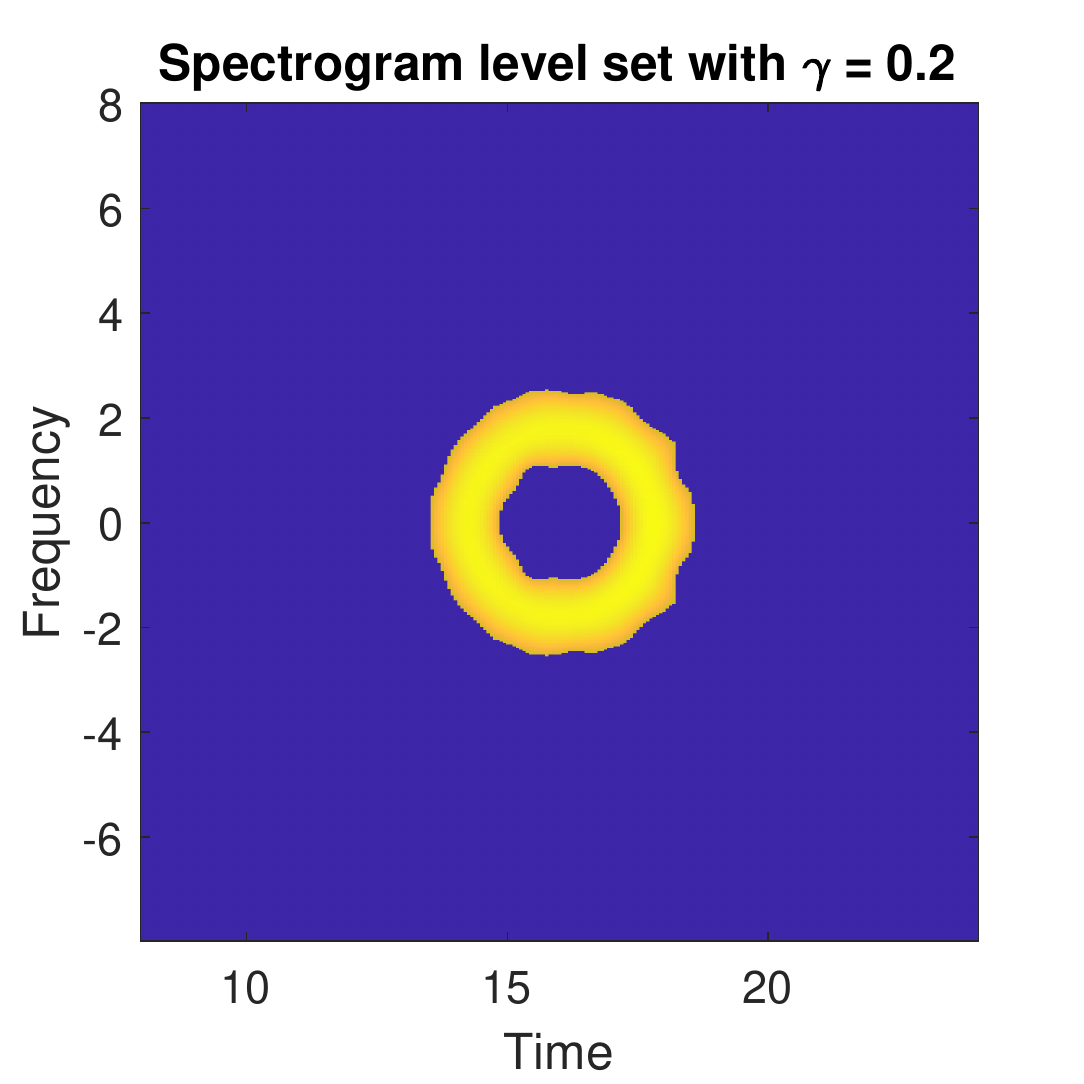}
	\vspace{-0.3cm}
	\caption{Gabor spectrogram and spectrogram level set of one fundamental mode corrupted by noise with $k=10$ in \eqref{eq: one_mode}.}\label{fig: Gabor_spectrogram_10}
\end{figure}

\begin{figure}[H]
	\centering
	\includegraphics[width=0.35\linewidth]{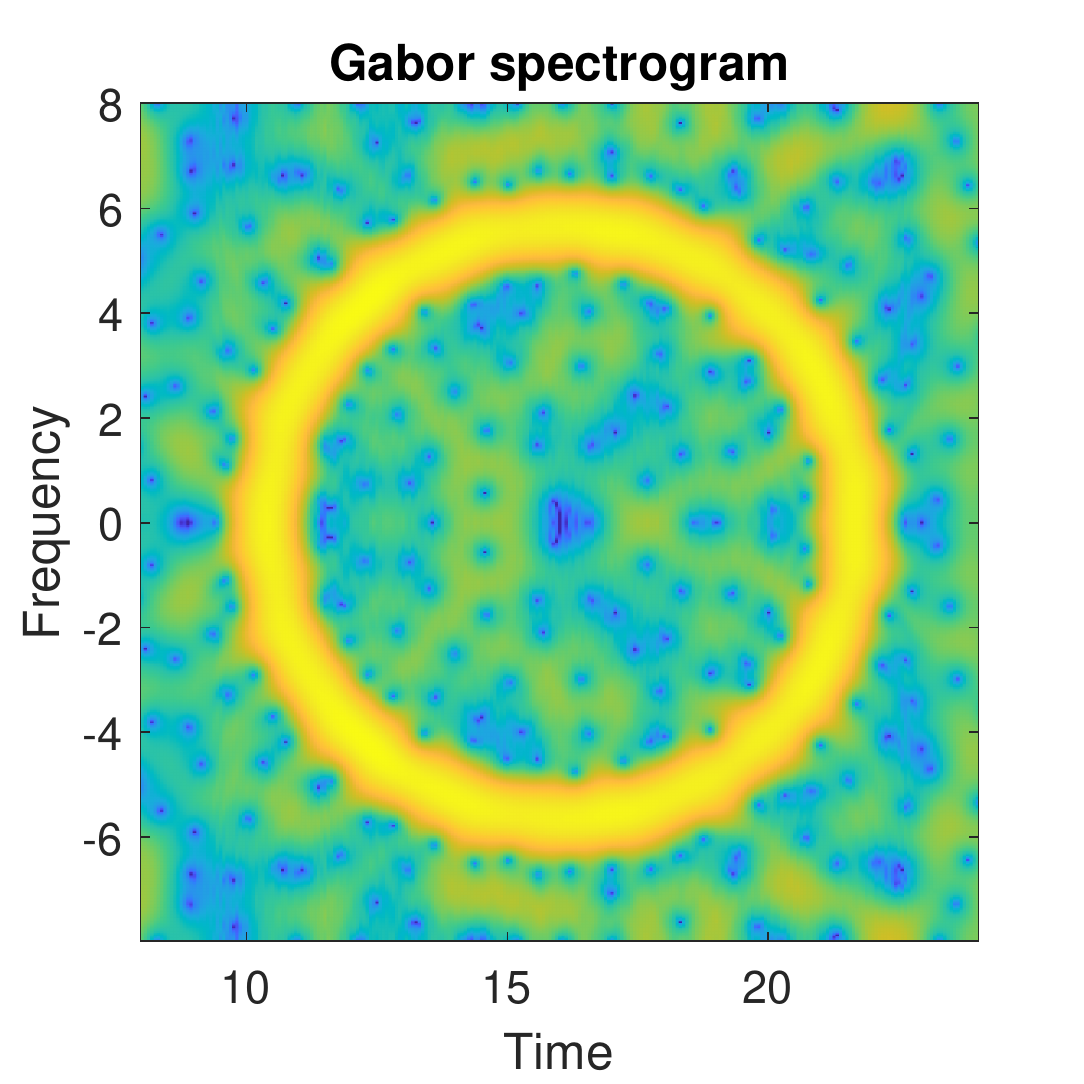}
	\includegraphics[width=0.35\linewidth]{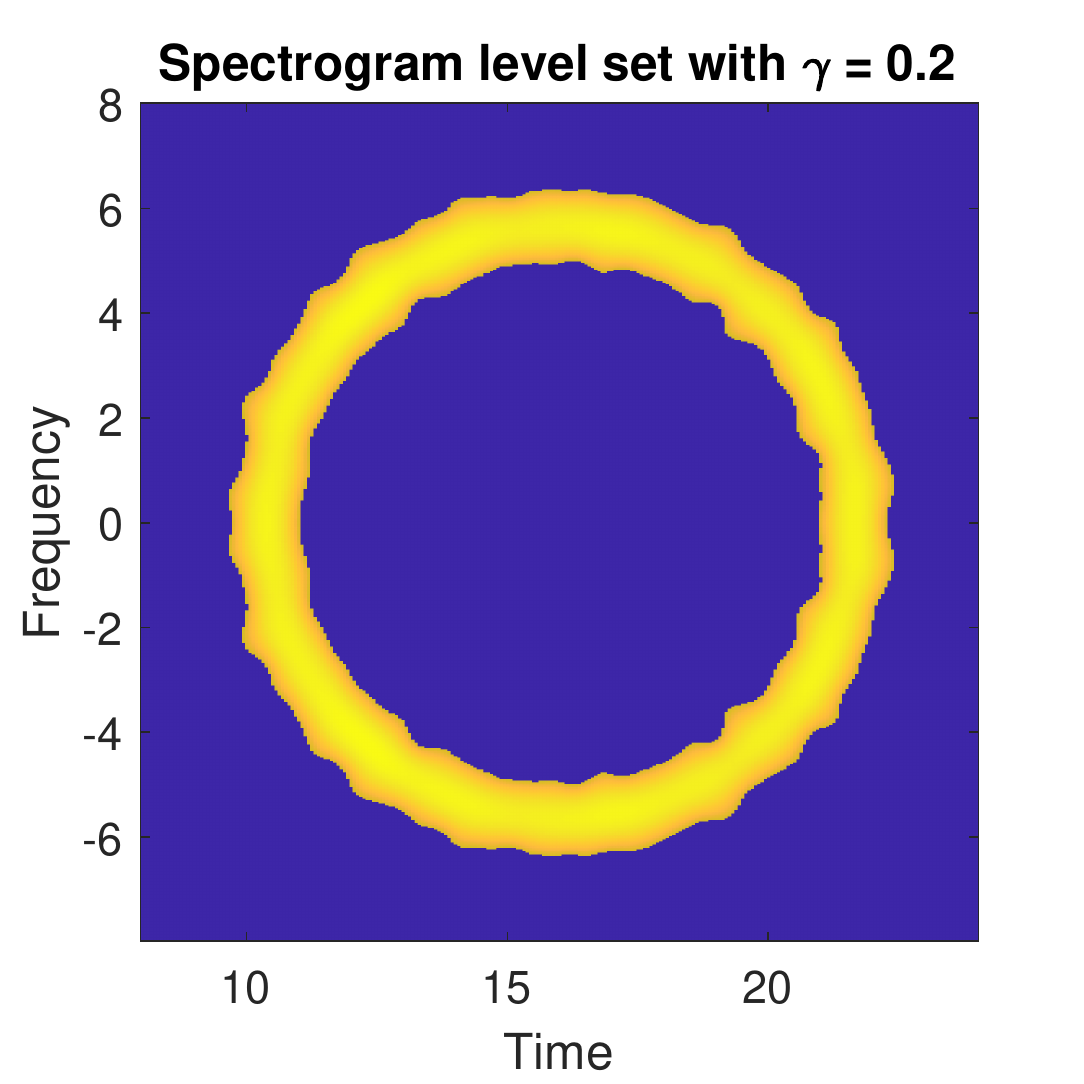}
	\vspace{-0.3cm}
	\caption{Gabor spectrogram and spectrogram level set of one fundamental mode corrupted by noise with $k=100$ in \eqref{eq: one_mode}.}\label{fig: Gabor_spectrogram_100}
\end{figure}

Figure \ref{fig: Gabor_spectrogram_10} and Figure \ref{fig: Gabor_spectrogram_100} show two examples to illustrate the Gabor spectrogram of data $y$ and its corresponding spectrogram level sets, where we fix $k=10$ and $k=100$, respectively. As we can see, it is promising to detect the fundamental mode through the annulus in the spectrogram level set. 

\vspace{-0.3cm}

The first row of Table \ref{table_one_mode} shows the empirical performance of Algorithm \ref{alg: one_signal} for learning one fundamental mode over $10000$ trails. One can see that the average mACC is 99.32\%, which implies that almost all trials could get the estimation in $\{k-1,k,k+1\}$ where $k$ is the true mode index.

\begin{table}[H]
	\renewcommand\arraystretch{1.1}
	\caption{Empirical performance of Algorithm \ref{alg: one_signal} for learning the linear combination of fundamental modes over $10000$ trails with different parameters.} \label{table_one_mode}
	\begin{center}
	\begin{tabular}{|c|c|c|}
		\hline
		Number of modes $(m)$ & The parameter $w$ & Average mACC  \\\hline
		1 & -- & 99.32\%  \\ 
		2 & 1.5 & 94.96\%  \\
		2 & 2  & 99.85\%  \\
		3 & 1.5 & 91.93\%  \\
		3 & 2 & 97.11\%  \\
		\hline
	\end{tabular}
\end{center}
\end{table}

\textbf{Two or more fundamental modes.} Note that Algorithm \ref{alg: one_signal} could be extended to estimate the linear combination of fundamental modes, as long as these modes are reasonably well separated. We then test the case of $m\geq 2$ fundamental modes, that is the generative model is given as
\begin{align}
y=\sum_{i=1}^m\lambda_i h_{k_i}+\sigma \xi,\label{eq: more_mode}
\end{align}
where signals $h_{k_i}$'s are Hermite functions defined in \eqref{eq: def_hk}, white noise $\xi$ is a random variable with distribution $\mu_1$, $\lambda_i$'s are the signal strength and $\sigma$ is the noise strength. In each trail, we 
\begin{itemize}
	\item take the observation radius $L=8$;
	\item randomly generate the integers $k_i$'s uniformly from $\{1,2,\cdots,112\}$ such that each $k_i$ are well separated as:
	\begin{align*}
	\sqrt{\frac{k_i}{\pi}}\notin [\sqrt{\frac{k_j}{\pi}}-w,\sqrt{\frac{k_j}{\pi}}+w],\quad 1\leq i\neq j \leq m,
	\end{align*}
	where $w>0$ is a preset parameter;
	\item randomly generate each parameter $\lambda_i$ uniformly in $[1,2]$;
	\item take the parameter $\sigma$ as $\frac{1}{10\sqrt{\log L}}$.
\end{itemize}
Note that in the generative model above, larger $w$ will give us a signal with more separate peaks, which makes it easier to detect and estimate the fundamental modes.

\begin{figure}[H]
	\centering
	\includegraphics[width=0.35\linewidth]{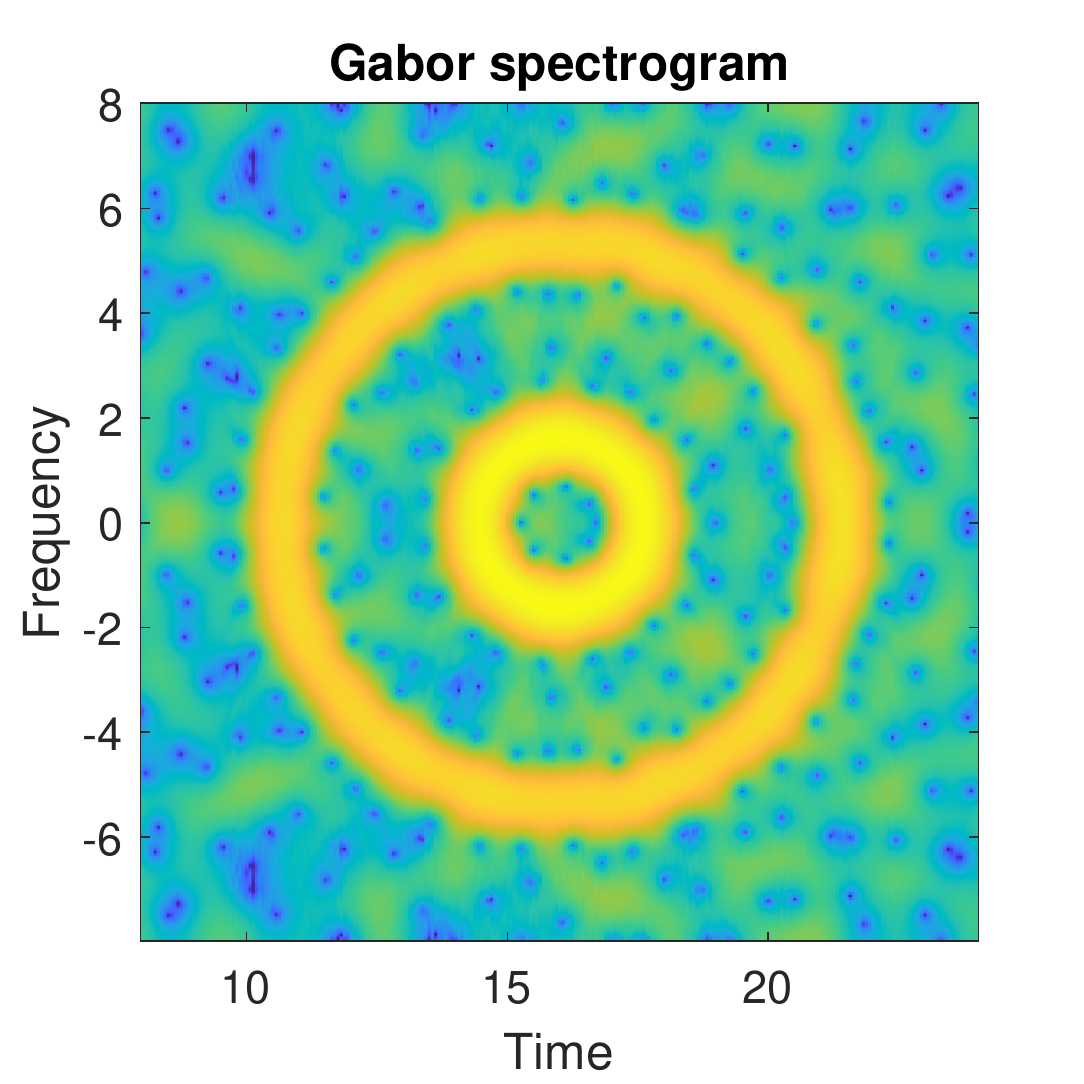}
	\includegraphics[width=0.35\linewidth]{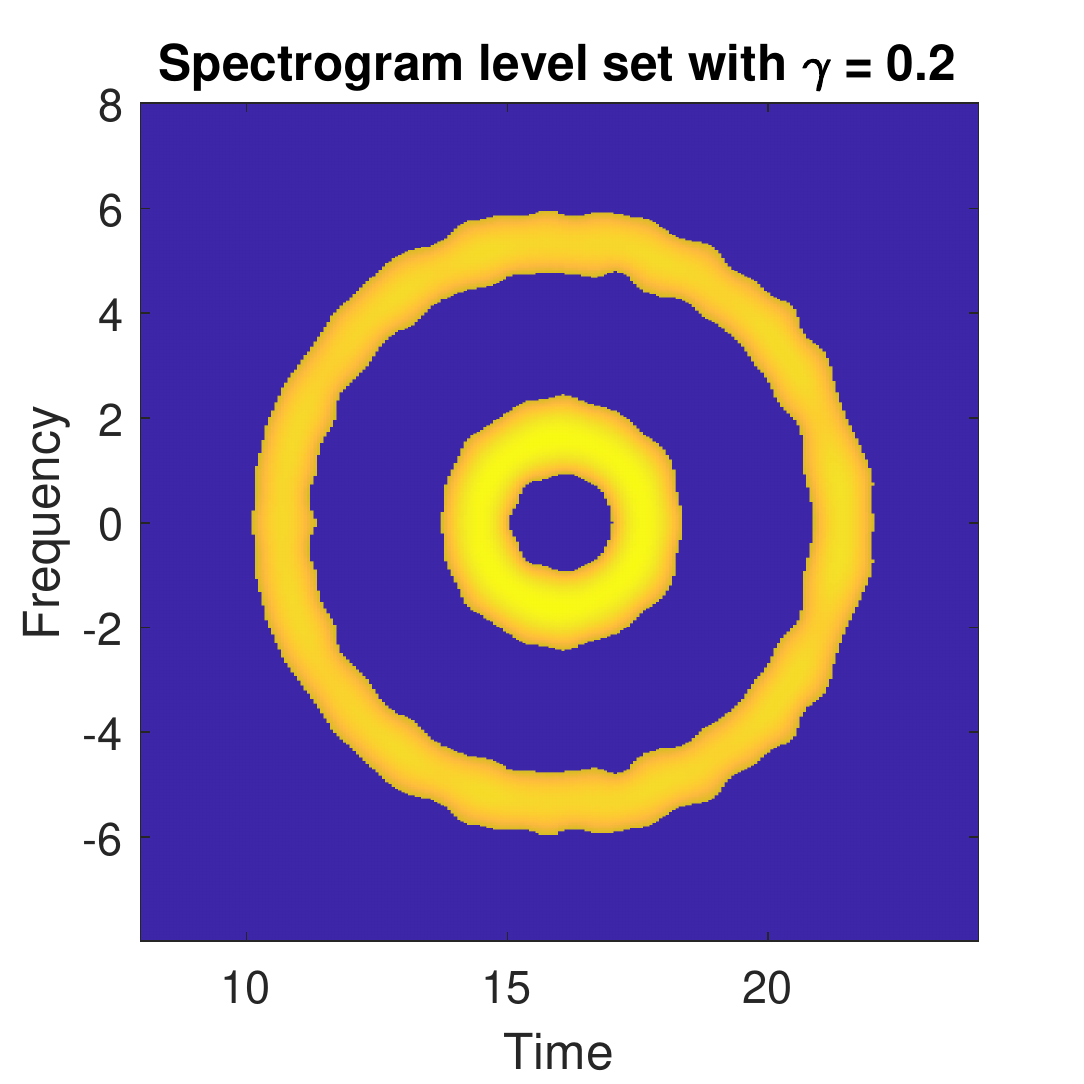}
	\vspace{-0.2cm}
	\caption{Gabor spectrogram and spectrogram level set of linear combination of fundamental modes corrupted by noise with $k_1=8$, $k_2=90$ in \eqref{eq: more_mode}.}\label{fig: Gabor_spectrogram_multiple_8_90}
\end{figure}

Figure \ref{fig: Gabor_spectrogram_multiple_8_90} and Figure \ref{fig: Gabor_spectrogram_multiple_10_40_95} show the Gabor spectrogram and spectrogram level set of data generated by two and three fundamental modes, respectively. It can be seen that, in these two cases, the annuli in the spectrogram level sets are well separated, which enables us to detect and estimate the signals via analysing the spectrogram level sets.

\vspace{-0.2cm}

Table \ref{table_one_mode} also shows the empirical performance of Algorithm \ref{alg: one_signal} for learning two or more fundamental modes over $10000$ trails with different choices of $w$. As we can see from the table, when the parameter $w$ is taken as the value of $2$, our algorithm gives more accurate estimation due to the fact that in this case, the fundamental modes are more separated. Overall, our algorithm is quite effective for learning linear combinations of fundamental modes as long as these modes are reasonably well separated.

\section{Proofs of main results}
In this section, we present the proofs of the main theoretical results in this paper. The proofs will be divided into different subsections, one pertaining to each result.

\subsection{Proof of Proposition \ref{prop: covariance}} \label{sec:covariance}

\begin{proof}
	Notice that $\mathbb{E}\left[{\cal F}[\xi](z)\right]=0$ for all $z\in \mathbb{C}$. Denote $\xi_k:=\langle \xi,h_k\rangle$, which is real valued as $\xi$ follows the white noise measure $\mu_1$. Then we can see that
	\begin{align*}
	& {\cal F}[\xi](z)\overline{{\cal F}[\xi](w)} \\ 
	= & \pi \exp(i\pi u_1v_1-i\pi u_2v_2-\frac{\pi}{2}|z|^2-\frac{\pi}{2}|w|^2) \times\sum_{k=0}^{\infty} \left[ \sum_{l=0}^k \left( \xi_l\frac{\pi^{l/2} z^l}{\sqrt{l!}}\right)
	\left( \xi_{k-l}\frac{\pi^{(k-l)/2} \overline{w}^{k-l}}{\sqrt{(k-l)!}}\right)\right].
	\end{align*}
	Therefore, due to the fact that $\{\xi_k\}$ are i.i.d. real standard Gaussian random variables, it holds that
	\begin{align*}
	&\mathbb{E}\left[{\cal F}[\xi](z)\overline{{\cal F}[\xi](w)}\right]\\
	&=\pi \exp(i\pi u_1v_1-i\pi u_2v_2-\frac{\pi}{2}|z|^2-\frac{\pi}{2}|w|^2)\times \sum_{k=0}^{\infty} \mathbb{E}\left[ \sum_{l=0}^k \left( \xi_l\frac{\pi^{l/2} z^l}{\sqrt{l!}}\right)
	\left( \xi_{k-l}\frac{\pi^{(k-l)/2} \overline{w}^{k-l}}{\sqrt{(k-l)!}}\right)\right]\\
	&=\pi \exp(i\pi u_1v_1-i\pi u_2v_2-\frac{\pi}{2}|z|^2-\frac{\pi}{2}|w|^2) \sum_{k=0}^{\infty}\frac{\pi^{k} z^k \overline{w}^{k}}{k!}\\
	&=\pi \exp(i\pi u_1v_1-i\pi u_2v_2-\frac{\pi}{2}|z|^2-\frac{\pi}{2}|w|^2+\pi z\overline{w}) .
	\end{align*}
	Thus by definition, the covariance function could be computed as
	\begin{align*}
	K(z,w) &=\mathbb{E}\left[{\cal F}[\xi](z)\overline{{\cal F}[\xi](w)}\right]\\
	&=\pi \exp(i\pi u_1v_1-i\pi u_2v_2-\frac{\pi}{2}|z|^2-\frac{\pi}{2}|w|^2+\pi z\overline{w}).
	\end{align*}
	Taking $w=z$ in the above formula, we have that
	\begin{align*}
	\sigma^2(z) =\mathbb{E}\left[|{\cal F}[\xi](z)|^2\right]=\pi e^{-\pi|z|^2} e^{\pi|z|^2} = \pi,
	\end{align*}
	which completes the proof.
\end{proof}

\subsection{Proof of Proposition \ref{prop: d}}\label{sec: proof_d}
\begin{proof}
	According to the definition of $d(\cdot,\cdot)$, we have that
	\begin{align*}
	d^2(z,w)&=\mathbb{E}\left[\left(F[\xi](z)-F[\xi](w)\right)\overline{\left(F[\xi](z)-F[\xi](w)\right)}\right]\\
	&=\sigma^2(z)+\sigma^2(w)-K(z,w)-K(w,z)\\
	&=2\pi-K(z,w)-\overline{K(z,w)}.
	\end{align*}
	Writing $z=u_1+iv_1$ and $w=u_2+iv_2$, we have that
	\begin{align*}
	K(z,w)+\overline{K(z,w)}&=2{\rm Re}\left(K(z,w)\right)\\
	&=2\pi \exp(-\frac{\pi}{2}|z|^2-\frac{\pi}{2}|w|^2+{\rm Re}\left(\pi z\overline{w}\right)) \times \cos\left(\pi u_1v_1-\pi u_2v_2 +{\rm Im}\left(\pi 	z\overline{w}\right)\right)\\
	&=2\pi \exp(-\frac{\pi}{2}|z-w|^2)\cos\left(\pi u_1v_1-\pi u_2v_2 -\pi u_1v_2 +\pi v_1 u_2\right)\\
	&=2\pi \cos\left(\pi(u_1+u_2)(v_1-v_2)\right)\exp(-\frac{\pi}{2}|z-w|^2).
	\end{align*}
	Therefore,
	\begin{align*}
	d^2(z,w)=2\pi \left[1-\cos  \left(\pi(u_1+u_2)(v_1-v_2) \right)\exp(-\frac{\pi}{2}|z-w|^2)\right],
	\end{align*}
	which completes the proof.
\end{proof}

\subsection{Proof of Theorem \ref{thm: boundedness_white_noise}} \label{sec: boundedness_white_noise}

\begin{proof}
	For any $z\in \mathbb{C}$, we could write
	\begin{align*}
	{\cal F}[\xi](z)={\cal F}_{\rm R}[\xi](z)+i{\cal F}_{\rm I}[\xi](z),
	\end{align*}
	where ${\cal F}_{\rm R}[\xi](z)$, ${\cal F}_{\rm I}[\xi](z)$ are real valued. Thus $\{{\cal F}_{\rm R}[\xi](z)\}_{z\in \mathbb{C}}$, $\{{\cal F}_{\rm I}[\xi](z)\}_{z\in \mathbb{C}}$ are (real) centered Gaussian random fields. 
	
	Due to Borell-TIS inequality \cite[Theorem 2.1.1]{adler2009random}, we have that for all $\rho>0$,
	\begin{align*}
	\mathbb{P}\left[\sup_{z\in \mathbb{B}_L} {\cal F}_{\rm R}[\xi](z)-\mathbb{E}\left[ \sup_{z\in \mathbb{B}_L} {\cal F}_{\rm R}[\xi](z)\right]>\rho
	\right] 
	\leq & \exp(-\frac{\rho^2}{2\sup_{z\in \mathbb{B}_L}\mathbb{E}\left[{\cal F}_{\rm R}[\xi](z)^2\right]}) \\
	\leq & \exp(-\frac{\rho^2}{2\sup_{z\in \mathbb{B}_L}\mathbb{E}\left[|{\cal F}[\xi](z)|^2\right]}) 
	=  \exp(-\frac{\rho^2}{2\pi}),
	\end{align*}
	where the last equality follows from Proposition \ref{prop: covariance}. As a result, with probability $\geq 1-\exp(-\frac{\rho^2}{2\pi})$, we have
	\begin{align*}
	&\sup_{z\in \mathbb{B}_L} {\cal F}_{\rm R}[\xi](z) \\
	&= \mathbb{E}\left[ \sup_{z\in \mathbb{B}_L} {\cal F}_{\rm R}[\xi](z)\right]+\left(\sup_{z\in \mathbb{B}_L} {\cal F}_{\rm R}[\xi](z)-\mathbb{E}\left[ \sup_{z\in \mathbb{B}_L} {\cal F}_{\rm R}[\xi](z)\right]\right)\\
	&\leq \mathbb{E}\left[ \sup_{z\in \mathbb{B}_L} {\cal F}_{\rm R}[\xi](z)\right]+\rho.
	\end{align*}
	According to \cite[Theorem 1.3.3]{adler2009random}, we know that there exists a universal
	constant $K_{\rm R}>0$ such that
	\begin{align*}
	\mathbb{E}\left[ \sup_{z\in \mathbb{B}_L} {\cal F}_{\rm R}[\xi](z)\right]\leq K_{\rm R}\int_{0}^{\infty} \sqrt{\log\left(N(\mathbb{B}_L,d_{\rm R},\varepsilon)\right)}\mathrm{d}\varepsilon,
	\end{align*}
	where $N(\mathbb{B}_L,d_{\rm R},\varepsilon)$ denotes the metric entropy for $\{{\cal F}_{\rm R}[\xi](z)\}_{z\in \mathbb{B}_L}$ with the associated canonical metric defined as
	$d_{\rm R}(z,w)=\left\{ \mathbb{E}\left[|F_{\rm R}[\xi](z)-F_{\rm R}[\xi](w)|^2\right]\right\}^{1/2}$. Thus it can be seen that for any $z,w\in \mathbb{B}_L$, $d_{\rm R}(z,w)\leq d(z,w)$, which indicates that for any $\varepsilon>0$,
	\begin{align*}
	N(\mathbb{B}_L,d_{\rm R},\varepsilon)\leq N(\mathbb{B}_L,d,\varepsilon).
	\end{align*}
	Therefore, with probability $\geq 1-\exp(-\frac{\rho^2}{2\pi})$, we have
	\begin{align}\label{eq: FR}
	\sup_{z\in \mathbb{B}_L} {\cal F}_{\rm R}[\xi](z)\leq K_{\rm R}\int_{0}^{\infty} \sqrt{\log\left(N(\mathbb{B}_L,d,\varepsilon)\right)}\mathrm{d}\varepsilon+\rho.
	\end{align}
	Similarly, we could prove that with probability $\geq 1-\exp(-\frac{\rho^2}{2\pi})$, 
	\begin{align}\label{eq: FI}
	\sup_{z\in \mathbb{B}_L} {\cal F}_{\rm I}[\xi](z) \leq K_{\rm I}\int_{0}^{\infty} \sqrt{\log\left(N(\mathbb{B}_L,d,\varepsilon)\right)}\mathrm{d}\varepsilon+\rho,
	\end{align}
	where $K_{\rm I}>0$ is a constant.
	
	Note that by the definition of $N(\mathbb{B}_L,d,\varepsilon)$, when 
	\begin{align*}
	\varepsilon\geq {\rm diam}(\mathbb{B}_L):=\sup_{z,w\in\mathbb{B}_L}d(z,w),
	\end{align*}
	we have $N(\mathbb{B}_L,d,\varepsilon)=1$. Moreover, by Proposition \ref{prop: d}, we can see that $d^2(z,w)\leq 4\pi$ for any $z,w\in \mathbb{C}$, which implies that ${\rm diam}(\mathbb{B}_L)\leq  2\sqrt{\pi}$. Thus
	\begin{align*}
	\int_{0}^{\infty} \sqrt{\log\left(N(\mathbb{B}_L,d,\varepsilon)\right)}\mathrm{d}\varepsilon
	&=\int_{0}^{2\sqrt{\pi}} \sqrt{\log\left(N(\mathbb{B}_L,d,\varepsilon)\right)}\mathrm{d}\varepsilon\\
	&=\int_{0}^{L^{-2}} \sqrt{\log\left(N(\mathbb{B}_L,d,\varepsilon)\right)}\mathrm{d}\varepsilon+\int_{L^{-2}}^{2\sqrt{\pi}}\sqrt{\log\left(N(\mathbb{B}_L,d,\varepsilon)\right)}\mathrm{d}\varepsilon.
	\end{align*}
	
	Next we will prove that when $0\leq \varepsilon\leq L^{-2}$, given $z,w\in \mathbb{B}_L$ with $z=u_1+iv_1$, $w=u_2+iv_2$, if $\max\{|u_1-u_2|,|v_1-v_2|\}\leq \frac{\varepsilon}{3\sqrt{2\pi^3}L}$, it holds that $d(z,w)\leq \varepsilon$. Suppose $z,w\in \mathbb{B}_L$ satisfies $\max\{|u_1-u_2|,|v_1-v_2|\}\leq \frac{\varepsilon}{3\sqrt{2\pi^3}L}$. It can be observed that for any $x\in \mathbb{R}$, $e^{-x}\geq 1-x$ and $\cos(x)\geq 1-x^2$. Based on these, we can see that
	\begin{align*}
	|\pi(u_1+u_2)(v_1-v_2)|\leq \frac{2\pi L \varepsilon}{3\sqrt{2\pi^3}L}\leq \frac{2\varepsilon}{3\sqrt{2\pi}}\leq \frac{2}{3\sqrt{2\pi}L^2}<1,
	\end{align*}
	which further implies
	\begin{align*}
	\cos  \left(\pi(u_1+u_2)(v_1-v_2) \right)>0.
	\end{align*}
	Thus, together with Proposition \ref{prop: d}, it holds that
	\begin{align*}
	d^2(z,w)&\leq 2\pi \left[1- \left(1-\pi^2(u_1+u_2)^2(v_1-v_2)^2 \right)\left( 1-\frac{\pi}{2}|z-w|^2\right)\right]\\
	&\leq 2\pi \left[\pi^2(u_1+u_2)^2(v_1-v_2)^2 +\frac{\pi}{2}|z-w|^2\right]\\
	&\leq (8\pi^3 L^2+\pi^2) |z-w|^2\leq 9\pi^3 L^2 \frac{\varepsilon^2}{9\pi^3L^2}=\varepsilon^2,
	\end{align*}
	where the last inequality follows from the fact that $L\geq \pi$, which means that $d(z,w)\leq \varepsilon$. 
	
	Due to the above discussion, when $0\leq \varepsilon\leq L^{-2}$, if $\{B_{d_{\rm E}}(z_i,\frac{\varepsilon}{3\sqrt{2\pi^3}L})\}_{i=1}^{M}$ is a covering of $\mathbb{B}_L$, where $d_{\rm E}(z,w):=\max\{|u_1-u_2|,|v_1-v_2|\}$, $\{B_{d}(z_i,\varepsilon)\}_{i=1}^{M}$ is also a covering of $\mathbb{B}_L$. Therefore, we have
	\begin{align*}
	N(\mathbb{B}_L,d,\varepsilon)\leq \left(\frac{2L}{\frac{2\varepsilon}{3\sqrt{2\pi^3}L}}\right)^2 =18\pi^3\frac{L^4}{\varepsilon^2},\quad 0\leq \varepsilon\leq L^{-2}.
	\end{align*}
	As a result, it holds that
	\begin{align*}
	\int_{L^{-2}}^{2\sqrt{\pi}}\sqrt{\log\left(N(\mathbb{B}_L,d,\varepsilon)\right)}\mathrm{d}\varepsilon
	&\leq 2\sqrt{\pi} \sqrt{\log\left(N(\mathbb{B}_L,d,L^{-2})\right)}\\
	&\leq 2\sqrt{\pi} \sqrt{\log\left(18\pi^3 L^8\right)}\\
	&\leq 2\sqrt{\pi} \sqrt{\log 2+2\log 3+3\log \pi +8\log L}\leq 2\sqrt{14\pi}\sqrt{\log L},
	\end{align*}
	and
	\begin{align*}
	\int_{0}^{L^{-2}} \sqrt{\log\left(N(\mathbb{B}_L,d,\varepsilon)\right)}\mathrm{d}\varepsilon &\leq \int_{0}^{L^{-2}} \sqrt{\log \left(18\pi^3\frac{L^4}{\varepsilon^2}\right)}\mathrm{d}\varepsilon
	\leq \int_{0}^{L^{-2}} \sqrt{\log \left(\frac{L^{10}}{\varepsilon^2}\right)}\mathrm{d}\varepsilon\\
	&\leq \sqrt{7}\int_{0}^{L^{-2}} \sqrt{\log \left(\frac{1}{\varepsilon}\right)}\mathrm{d}\varepsilon
	=\sqrt{7}\left(\frac{\sqrt{\log L}}{L^2}+\int_{\sqrt{2\log L}}^{\infty} e^{-t^2}\mathrm{d}t\right).
	\end{align*}
	According to Mills’ ratio inequality \cite{gordon1941values}, we know that
	\begin{align*}
	e^{\frac{x^2}{2}}\int_{x}^{\infty} e^{-\frac{t^2}{2}}\mathrm{d}t<\frac{1}{x},
	\end{align*}
	which means
	\begin{align*}
	\int_{\sqrt{2\log L}}^{\infty} e^{-t^2}\mathrm{d}t<\frac{1}{2\sqrt{2}L^2\sqrt{\log L}}.
	\end{align*}
	Therefore, it holds that
	\begin{align*}
	\int_{0}^{L^{-2}} \sqrt{\log\left(N(\mathbb{B}_L,d,\varepsilon)\right)}\mathrm{d}\varepsilon& \leq \sqrt{7}\left(\frac{\sqrt{\log L}}{L^2}+\frac{1}{2\sqrt{2}L^2\sqrt{\log L}}\right)\\
	&<\frac{4\sqrt{7}+\sqrt{14}}{4\pi^2}\sqrt{\log L},
	\end{align*}
	which further implies
	\begin{align*}
	\int_{0}^{\infty} \sqrt{\log\left(N(\mathbb{B}_L,d,\varepsilon)\right)}\mathrm{d}\varepsilon
	&\leq (2\sqrt{14\pi}+\frac{4\sqrt{7}+\sqrt{14}}{4\pi^2})\sqrt{\log L}\\
	&<14\sqrt{\log L}.
	\end{align*}
	
	Take $\rho = \tau\sqrt{\log L}$ in \eqref{eq: FR}, we have with probability  $\geq 1-\exp\left(-\frac{\tau^2}{2\pi} \cdot \log L\right)$,
	\begin{align*}
	\sup_{z\in \mathbb{B}_L} {\cal F}_{\rm R}[\xi](z)&\leq K_{\rm R}\int_{0}^{\infty} \sqrt{\log\left(N(\mathbb{B}_L,d,\varepsilon)\right)}\mathrm{d}\varepsilon+\tau\sqrt{\log L}\\
	&\leq (14K_{\rm R}+\tau)\sqrt{\log L}.
	\end{align*}
	Similarly, according to \eqref{eq: FI}, we have with probability  $\geq 1-\exp\left(-\frac{\tau^2}{2\pi} \cdot \log L\right)$,
	\begin{align*}
	\sup_{z\in \mathbb{B}_L} {\cal F}_{\rm I}[\xi](z)\leq (14K_{\rm I}+\tau)\sqrt{\log L}.
	\end{align*}
	Denote $K=\max\{K_R, K_I\}$, then it holds that
	\begin{align*}
	&\mathbb{P}\left[ \sup_{z\in \mathbb{B}_L}|{\cal F}[\xi](z)|> \sqrt{2}(14K+\tau)\sqrt{\log L}\right]\\
	&\leq\mathbb{P}\left[ \sup_{z\in \mathbb{B}_L}|{\cal F}_{\rm R}[\xi](z)|> (14K+\tau)\sqrt{\log L}\right]+\mathbb{P}\left[ \sup_{z\in \mathbb{B}_L}|{\cal F}_{\rm I}[\xi](z)|> (14K+\tau)\sqrt{\log L}\right]\\
	&\leq2\mathbb{P}\left[ \sup_{z\in \mathbb{B}_L}{\cal F}_{\rm R}[\xi](z)> (14K+\tau)\sqrt{\log L}\right]+2\mathbb{P}\left[ \sup_{z\in \mathbb{B}_L}{\cal F}_{\rm I}[\xi](z)> (14K+\tau)\sqrt{\log L}\right],
	\end{align*}
	where the last inequality follows from symmetry. As a result, we have
	\begin{align*}
	&\mathbb{P}\left[ \sup_{z\in \mathbb{B}_L}|{\cal F}[\xi](z)|\leq \sqrt{2}(14K+\tau)\sqrt{\log L}\right]\\
	&\geq 2\mathbb{P}\left[ \sup_{z\in \mathbb{B}_L}{\cal F}_{\rm R}[\xi](z)\leq (14K+\tau)\sqrt{\log L}\right]+2\mathbb{P}\left[ \sup_{z\in \mathbb{B}_L}{\cal F}_{\rm I}[\xi](z)\leq (14K+\tau)\sqrt{\log L}\right]-3\\
	&\geq 2\mathbb{P}\left[ \sup_{z\in \mathbb{B}_L}{\cal F}_{\rm R}[\xi](z)\leq (14K_{\rm R}+\tau)\sqrt{\log L}\right]+2\mathbb{P}\left[ \sup_{z\in \mathbb{B}_L}{\cal F}_{\rm I}[\xi](z)\leq (14K_{\rm I}+\tau)\sqrt{\log L}\right]-3\\
	&\geq1-4\exp\left(-\frac{\tau^2}{2\pi} \cdot \log L\right),
	\end{align*}
	which completes the proof.
\end{proof}

\subsection{Proof of Proposition \ref{prop:max_spec_signal}} \label{sec:max_spec_signal}

\begin{proof}
	According to Proposition \ref{prop: stft_hk}, we have for any $u,v\in \mathbb{R}$,
	\begin{align}
	|V_g h_k (u,v)| = \exp(-\frac{\pi}{2}|z|^2)\frac{\pi^{k/2}|z|^k}{\sqrt{k!}},\label{eq: magnitude_Vghk}
	\end{align}
	where $z\in \mathbb{C}$ is taken as $z=u+iv$. Define $f:\mathbb{R}\rightarrow \mathbb{R}$ as
	$
	f(x) = e^{-\frac{\pi}{2}x^2}\frac{\pi^{k/2}x^k}{\sqrt{k!}}.
	$
	Due to the fact that
	\begin{align*}
	f'(x) = e^{-\frac{\pi}{2}x^2}\frac{\pi^{k/2}x^{k-1}}{\sqrt{k!}}\left(-\pi x^2+k \right),
	\end{align*}
	we know for any $x\geq 0$,
	\begin{align*}
	f(x)\leq f(\sqrt{\frac{k}{\pi}})=\prod_{t=1}^k\sqrt{\frac{k}{et}}.
	\end{align*}
	Therefore, together with the equation \eqref{eq: magnitude_Vghk}, we have that 
	\begin{align*}
	\max_{u,v\in \mathbb{R}}|V_g h_k (u,v)| = \prod_{t=1}^k \sqrt{\frac{k}{et}},
	\end{align*}
	where the maximum is obtained when $\sqrt{u^2+v^2}=\sqrt{k/\pi}$.
	
	In addition, for any $u,v\in \mathbb{R}$ such that $\sqrt{u^2+v^2}=(1+r)\sqrt{k/\pi}$ with $r\in \mathbb{R}$, it holds that
	\begin{align*}
	\frac{|V_g h_k (u,v)|}{\max_{u,v\in \mathbb{R}}|V_g h_k (u,v)|}&=\frac{e^{-\frac{k}{2}(1+r)^2}(1+r)^k(\sqrt{k/\pi})^k}{e^{-\frac{k}{2}}(\sqrt{k/\pi})^k}=\frac{(1+r)^k}{e^{k(r+r^2/2)}},
	\end{align*}
	which completes the proof.
\end{proof}

\subsection{Proof of Theorem \ref{thm:stoch-geom-signal}} \label{sec:stoch-geom-signal}
\begin{proof}
	By the definition of the STFT in \eqref{eq: def_stft}, we can see that
	\begin{align*}
	V_g y(u,v) = \lambda V_g h_k(u,v)+ V_g \xi(u,v),\quad  u,v\in \mathbb{R}.
	\end{align*}
	Thus it holds that for any $u,v\in \mathbb{R}$,
	\begin{align*}
	|\lambda||V_g h_k(u,v)|+|{\cal F}[\xi](z)|&\geq |V_g y(u,v)| \geq  |\lambda||V_g h_k(u,v)|-|{\cal F}[\xi](z)|,
	\end{align*}
	where $z=u+iv\in \mathbb{C}$ and ${\cal F}[\xi](z)=V_g \xi(u,v)$ as in \eqref{eq: def_Fxi}. According to Theorem \ref{thm: boundedness_white_noise}, for $L\geq\pi$ and $\tau>0$, we have 
	\begin{align*}
	\mathbb{P}\left[ \sup_{z\in \mathbb{B}_L}|{\cal F}[\xi](z)|\leq \sqrt{2}(14K+\tau)\sqrt{\log L}\right]
	\geq 1-4\exp\left(-\frac{\tau^2}{2\pi} \cdot \log L\right),
	\end{align*}
	which means
	\begin{align*}
	\sup_{z\in \mathbb{B}_L}\bigg||V_g y(u,v)|-|\lambda||V_g h_k(u,v)|\bigg| \leq \sqrt{2}(14K+\tau)\sqrt{\log L}
	\end{align*}
	with probability $\geq 1-4\exp\left(-\frac{\tau^2}{2\pi} \cdot \log L\right)$. We then consider the case when the above inequality holds.
	
	We first prove that the level set $\Lambda\left(3\sqrt{2}(14K+\tau)\sqrt{\log L}\right)\neq \emptyset$. Consider $u_0+i v_0\in \mathbb{B}_L$ such that $\sqrt{u_0^2+v_0^2}=\sqrt{k/\pi}$, we have
	\begin{align*}
	|V_g y(u_0,v_0)|&\geq |\lambda||V_g h_k(u_0,v_0)|-\bigg||V_g y(u_0,v_0)|-|\lambda||V_g h_k(u_0,v_0)|\bigg|\\
	&\geq |\lambda|\prod_{t=1}^k \sqrt{\frac{k}{et}}-\sup_{u+iv\in \mathbb{B}_L}\bigg||V_g y(u,v)|-|\lambda||V_g h_k(u,v)|\bigg|\\
	&\geq 5\sqrt{2}(14K+\tau)\sqrt{\log L}-\sqrt{2}(14K+\tau)\sqrt{\log L}\\
	&=4 \sqrt{2}(14K+\tau)\sqrt{\log L}.
	\end{align*}
	That is to say,
	\begin{align*}
	(u_0,v_0)\in \Lambda\left(3\sqrt{2}(14K+\tau)\sqrt{\log L}\right).
	\end{align*}
	
	Next we prove that 
	\begin{align*}
	\Lambda\left(3\sqrt{2}(14K+c)\sqrt{\log L}\right) \subseteq \left\{(u,v): \frac{|V_g h_k (u,v)|}{\prod_{t=1}^k \sqrt{k/(et)}}> \alpha\right\}.
	\end{align*}
	For any $u+iv\in \mathbb{B}_L$ such that $|V_g h_k (u,v)|\leq \alpha \prod_{t=1}^k \sqrt{k/(et)}$, we can see that
	\begin{align*}
	|V_g y (u,v)|&\leq |\lambda||V_g h_k (u,v)|+\sup_{u+iv\in \mathbb{B}_L}\bigg||V_g y(u,v)|-|\lambda||V_g h_k(u,v)|\bigg|\\
	&\leq |\lambda|\alpha \prod_{t=1}^k \sqrt{k/(et)}+\sqrt{2}(14K+\tau)\sqrt{\log L}\\
	&=2 \sqrt{2}(14K+\tau)\sqrt{\log L}.
	\end{align*}
	which indicates that
	\begin{align*}
	(u,v)\notin \Lambda\left(3\sqrt{2}(14K+\tau)\sqrt{\log L}\right).
	\end{align*}
	
	Therefore, from the above discussion, we have that 
	\begin{align*}
	\emptyset \neq \Lambda\left( 3\sqrt{2}(14K + \tau)\sqrt{\log L}\right) 
	\subseteq \left\{ (u,v): \frac{|V_g h_k (u,v)|}{\prod_{t=1}^k \sqrt{k/(et)}}> \alpha\right\}
	\end{align*}
	with probability $ \ge 1-4\exp\left(-\frac{\tau^2}{2\pi} \cdot \log L\right)$. 
\end{proof}

\subsection{Proof of Theorem \ref{thm:hyp-test}} \label{sec:hyp-test}
\begin{proof}
	We observe that if we set $\tau=\sqrt{2\pi \cdot \log (4/\del) / \log L}$, then the conclusion of Theorem \ref{thm:stoch-geom-signal} applies with probability $\ge 1 -\del$, the level set \[\L\l(3\sqrt{2}\l(14K \sqrt{\log L} + \sqrt{2\pi \cdot \log (4/\del)} \r)\r)\] and, for the fundamental signal $h_k$, the required signal strength
	\begin{align*} 
	|\la_k| \ge 5\sqrt{2} \l(\prod_{t=1}^k\sqrt{k/(et)}\r)^{-1}\l(14K \sqrt{\log L} + \sqrt{2\pi \log (4/\del)}\r).
	\end{align*} 
	In addition, according to \eqref{eq:minimax} and \eqref{eq: signal_strengh_detection}, we observe that the choice of signal strength in the statement of this theorem satisfies 
	\begin{align*}
	|\la| \ge 5\sqrt{2}  \l(\prod_{t=1}^k\sqrt{k/(et)}\r)^{-1}  \l(14K \sqrt{\log L} + \sqrt{2\pi \cdot \log (4/\del)}\r)
	\end{align*} 
	for all $1 \le k \le k_0$.
	
	In view of the above discussion, we can conclude that, for the given choices of level set $\L(\t(\del))$ and signal strength $|\la|$ as in the statement of the present theorem, the conclusion of Theorem \ref{thm:stoch-geom-signal} holds with probability $\ge 1-\del$. Thus, if a fundamental signal $h_k$ is present (under the generative model as in Section \ref{sec:gen_obs_model}) with the prescribed signal strength, then $\P_k[\psi_{\t(\del)} = 0] \le \del$. 
	
	Further, with the same choices, the conclusion of Theorem \ref{thm: boundedness_white_noise} can also be verified to hold with probability $\ge 1-\del$. Thus, if no signal is present (under the generative model as in Section \ref{sec:gen_obs_model}), then $\P^0[\psi_{\t(\del)} = 1] \le \del$. 
	
	In consideration of this, and the definition of \textit{signal detection} as laid out in Section \ref{sec:signal detection}, we could conclude that the prescribed test as in Section \ref{sec:signal detection} performs signal detection at the  signal strength in the statement of the present theorem.
\end{proof}

\subsection{Proof of Theorem \ref{thm:estimation-k}} \label{sec:estimation-k}
\begin{proof}
	We consider the setting where $y=\la h_k + \xi$ for a fixed $k \in [1,\ldots,k_0]$. Let $\O_\tau$ denote the event
	\[ \O_\tau:=\{ \max_{u+iv\in \B_L} |V_g \xi (u,v)| \le \sqrt{2}(14K + \tau)\sqrt{\log L}\}. \] 
	From Theorem \ref{thm: boundedness_white_noise}, we have $\P[\O_\tau]\ge 1 - 4\exp\left(-\frac{\tau^2}{2\pi} \cdot \log L\right)$. Denote
	\begin{align*}
	{\cal M}_k&=\left\{u+iv\in \mathbb{B}_L:  (u,v) \in \arg\max_{u+iv\in\mathbb{B}_L} |V_g y (u,v)|^2\right\}.
	\end{align*}
	We {\bf claim} that on the event $\O_\tau$, for any $z \in \cM_k$, the quantity $\pi \cdot |z|^2$ satisfies 
	\[ k - \frac{1}{2} < \pi \cdot |z|^2 < k + \frac{1}{2},\]
	where the proof of the {\bf claim} is presented later. Thus, on the event $\O_\tau$,  for any $z \in \cM_k$ we have $[[\pi \cdot |z|^2]]=k$. Since $\P[\O_\tau] \ge 1 - 4\exp\left(-\frac{\tau^2}{2\pi} \cdot \log L\right)$, and the above analysis holds for every $1 \le k \le k_0$, we may deduce that  \[\inf_{1 \le k \le k_0}\P_{k}[\hk=k] \ge 1 - 4\exp\left(-\frac{\tau^2}{2\pi} \cdot \log L\right),  \] as desired.
	
	We now prove the {\bf claim}. We are going to prove that on the event $\O_\tau$, if $z \in \cM_k$, then it must hold that $\l|\pi \cdot |z|^2 - k \r| < 1/2$. This condition can be elaborated as 
	\[
	\l(1-\frac{1}{2k} \r)^{1/2} < \frac{|z|} {\sqrt{{k}/{\pi}}} < \l(1+\frac{1}{2k} \r)^{1/2}.
	\]
	This can further be expressed as
	\[
	-\frac{\sqrt{2k}-\sqrt{2k-1}}{\sqrt{2k}} < \frac{|z|} {\sqrt{{k}/{\pi}}} -1  < \frac{\sqrt{2k+1}-\sqrt{2k}}{\sqrt{2k}};
	\]
	equivalently,
	\begin{align} \label{eq:estimation-condition}
	-\frac{1}{\sqrt{2k}(\sqrt{2k}+\sqrt{2k-1})} < \frac{|z|} {\sqrt{{k}/{\pi}}} -1  < \frac{1}{\sqrt{2k}(\sqrt{2k+1}+\sqrt{2k})}.
	\end{align}
	We next define the set ${\cal A}_k$ as
	\begin{align*}
	{\cal A}_k:=\left\{z\in \mathbb{B}_L: -\frac{1}{4k+2}<\frac{|z|}{\sqrt{k/\pi}}-1<\frac{1}{4k+2}
	\right\}.
	\end{align*}
	Observe that any $z \in {\cal A}_k$ satisfies \eqref{eq:estimation-condition}. We will demonstrate that, in fact, on the event $\O_\tau$, $\cM_k \subset {\cal A}_k$. We prove this by deducing that $A_k^\c \subset \cM_k^\c$ on the event $\O_\tau$. Note that the complement of the set is taken with respect to the universal set $\mathbb{B}_L$. The proof is divided into three steps.
	
	\textbf{Step 1.} Suppose $z_0=u_0+iv_0\in {\cal A}_k^{\complement}$, which implies
	\begin{align*}
	\left|\frac{|z_0|}{\sqrt{k/\pi}}-1\right|\geq \frac{1}{4k+2}.
	\end{align*}
	We may invoke \eqref{eq: max_ratio} with $|r_0|=\l|\frac{|z_0|} {\sqrt{{k}/{\pi}}} -1\r|$ and the definition of $\beta(k)$ in \eqref{eq:beta} to conclude that 
	\begin{equation} \label{eq:decay}
	\frac{ |V_g h_k (u_0,v_0)|}{\max_{u,v\in \mathbb{R}} |V_g h_k (u,v)|} = C(k,r_0) \le  \beta(k).
	\end{equation}
	For $L\geq\max\{\sqrt{k_0/\pi},\pi\}$, according to Proposition \ref{prop:max_spec_signal}, we have
	\begin{align}
	\max_{u,v\in \mathbb{R}} |V_g h_k (u,v)| = \max_{u+iv\in \mathbb{B}_L} |V_g h_k (u,v)|.\label{eq: max_max}
	\end{align}
	Now, on the event $\O_\tau$ we have
	\begin{align*} 
	& |V_g y (u_0,v_0)| 
	\le  |\la| |V_g h_k (u_0,v_0)| + \max_{u+iv\in\mathbb{B}_L} |V_g \xi (u,v)| \\
	\le &  |\la| \max_{u+iv\in\mathbb{B}_L} |V_g h_k (u,v)| \ \beta(k)  +  \max_{u+iv\in\mathbb{B}_L} |V_g \xi (u,v)| \quad \text{ [using \eqref{eq:decay}, \eqref{eq: max_max}]}\\
	= &  |\la| \max_{u+iv\in\mathbb{B}_L}|V_g h_k (u,v)| \ \beta(k)  \l( 1 +\frac{ \max_{u+iv\in\mathbb{B}_L} |V_g \xi (u,v)|}{ |\la| \max_{u+iv\in\mathbb{B}_L} |V_g h_k (u,v)| \beta(k)}  \r) \\
	\le &  |\la| \max_{u+iv\in\mathbb{B}_L} |V_g h_k (u,v)| \ \beta(k) \l(1 + \frac{  1 }{ 5 C(k_0) \beta(k)}  \r) , 
	\end{align*}
	where, in the last step, we have use the threshold for $|\la|$ as in the statement of the present theorem, the definition of $\fM(k_0)$ and the event $\O_\tau$.  
	
	\textbf{Step 2.} Let ${\cal T}_k$ be the circle of radius $\sqrt{k/\pi}$ as: ${\cal T}_k:=\{z\in \mathbb{B}_L : |z| = \sqrt{k/\pi}\}$. For $z_1=u_1+iv_1 \in {\cal T}_k$, on the event $\O_\tau$ we have
	\begin{align*}  
	|V_g y (u_1,v_1)| 
	\ge& |\la|  |V_g h_k (u_1,v_1)| \l(1 -\frac{\max_{u+iv \in \B_L} |V_g \xi (u,v)| }{|\la|  |V_g h_k (u_1,v_1)|}\r) \\
	= &|\la|  \max_{u+iv\in\mathbb{B}_L}  |V_g h_k (u,v)| \l(1 - \frac{\max_{u+iv \in \B_L} |V_g \xi (u,v)| }{|\la|   \max_{u+iv\in\mathbb{B}_L}  |V_g h_k (u,v)|}\r) \\
	\ge  &|\la|  \max_{u+iv\in\mathbb{B}_L}  |V_g h_k (u,v)| \l(1 - \frac{1}{5 C(k_0)}\r),
	\end{align*}
	where, the equality in the second last step is due to Proposition \ref{prop:max_spec_signal} and in the last step, we have use the threshold for $|\la|$, the definition of $\fM(k_0)$ and the event $\O_\tau$.  
	
	\textbf{Step 3.} From Step 1 and Step 2, we have that
	\begin{align*}
	\sup_{u+iv \in {\cal A}_k^\c} |V_g y (u,v)| &\le   |\la| \max_{u+iv\in\mathbb{B}_L} |V_g h_k (u,v)| \beta(k) \l(1 + \frac{  1 }{ 5 C(k_0) \beta(k)}  \r)\\
	\inf_{u+iv\in {\cal T}_k} |V_g y (u,v)| &\geq |\la|  \max_{u+iv\in\mathbb{B}_L}  |V_g h_k (u,v)| \l(1 - \frac{1}{5 C(k_0)}\r)
	\end{align*}
	Now, we can invoke the definition of $C(k_0)$ as in \eqref{eq:cknot} and observe  that \[ \max_{1 \le k \le k_0} \beta(k) \cdot \l(1 + \frac{  1 }{ 5 C(k_0) \beta(k)}  \r) <  \l(1 - \frac{1}{5 C(k_0)}\r). \]
	This implies that $\inf_{u+iv\in {\cal T}_k} |V_g y (u,v)|  > \sup_{u+iv \in {\cal A}_k^\c} |V_g y (u,v)|$. Recalling the definition of $\cM_k$, we therefore deduce that $A_k^\c \subset \cM_k^\c$ on the event $\Omega_{\tau}$. This completes the proof.
\end{proof}

\subsection{Proof of Theorem \ref{thm:estimation-lambda}} \label{sec:estimation-lambda}
\begin{proof}
	We consider the setting where $y=\la h_k + \xi$ for a fixed $k \in [1,\ldots,k_0]$. Define $\tau(\del)=\del \sqrt{\log L} - 14K>0$.  Let $\O_{\tau(\del)}$ denote the event
	\[ \O_{\tau(\del)} :=\{ \max_{u+iv\in \B_L} |V_g \xi (u,v)| \le \sqrt{2}(14K + \tau(\del))\sqrt{\log L}\}. \] Using Theorem \ref{thm: boundedness_white_noise}, we may conclude that 
	\begin{align*}
	\P[\O_{\tau(\del)}]&\ge 1 - 4\exp\left(-\frac{\tau(\del)^2}{2\pi} \cdot \log L\right)= 1 - 4\exp\left(-\frac{\del^2 \log^2 L}{2\pi} \cdot \l(1-\frac{14K}{\del \sqrt{\log L}}\r)^2 \right).
	\end{align*}
	
	Using the fact that $0<\del \le 1$, we may deduce that the threshold $t_{\mathrm{strength}}$ as in the statement of the present theorem satisfies 
	\begin{align*} 
	t_{\mathrm{strength}} \ge  5 C(k_0) \sqrt{2}(14K+ \tau(\del))\fM(k_0)^{-1}\sqrt{\log L} = t_{\mathrm{mode}}(\tau(\del)).  
	\end{align*}
	Thus we may invoke Theorem \ref{thm:estimation-k} on the event $\O_{\tau(\del)}$ above and the signal strength $t_{\mathrm{strength}}$ in order to obtain $\hk=k$.
	
	For the rest of this proof, we will work on the event $\O_{\tau(\del)}$, whence $\hk=k$ as discussed above. For any $u_0+iv_0 \in \mathbb{B}_L$, we have
	\begin{align*}
	|\la| |V_g h_k (u_0,v_0)|& - \max_{u+iv\in\mathbb{B}_L} |V_g \xi (u,v)| \le |V_g y (u_0,v_0)| \le |\la| |V_g h_k (u_0,v_0)| + \max_{u+iv\in\mathbb{B}_L} |V_g \xi (u,v)|,
	\end{align*}
	which implies that 
	\begin{align*}
	|\la| \max_{u+iv\in \mathbb{B}_L}&|V_g h_k (u,v)| - \max_{u+iv\in \mathbb{B}_L} |V_g \xi (u,v)| \le \max_{u+iv\in \mathbb{B}_L} |V_g y (u,v)| \\
	&\le |\la| \max_{u+iv\in \mathbb{B}_L}|V_g h_k (u,v)| + \max_{u+iv\in \mathbb{B}_L} |V_g \xi (u,v)|.
	\end{align*}
	The above inequality could in turn be rewritten as:
	\begin{align*} 
	\l| \frac{\max_{u+iv\in \mathbb{B}_L} |V_g y (u,v)|}{|\la| \max_{u+iv\in \mathbb{B}_L}|V_g h_k (u,v)|} -1\r|\leq \frac{\max_{u+iv\in \mathbb{B}_L} |V_g \xi (u,v)|}{|\la| \max_{u+iv\in \mathbb{B}_L}|V_g h_k (u,v)| }. 
	\end{align*}
	Note that by the definition of $\hat{\la}$, we obtain 
	\begin{equation*} 
	\l| \frac{\hl}{|\la|} - 1 \r| \le \frac{\max_{u+iv\in \mathbb{B}_L} |V_g \xi (u,v)|}{|\la| \max_{u+iv\in \mathbb{B}_L}|V_g h_k (u,v)| }. 
	\end{equation*}
	Using the threshold $|\la|\ge 5C(k_0)\sqrt{2}\ \fM(k_0)^{-1} \log L$ as in the statement of the present theorem, we can see that on the event $\O_{\tau(\del)}$,
	\[ \l| \frac{\hl}{|\la|} - 1 \r| \le \frac{14K + \tau(\del)}{5C(k_0)\sqrt{\log L}}
	\leq \frac{14K + \tau(\del)}{\sqrt{\log L}}. \]
	Using $\tau(\del)=\del \sqrt{\log L} - 14K$, we obtain $\frac{14K + \tau(\del)}{\sqrt{\log L}}=\del$, implying $\l| \frac{\hl}{|\la|} - 1 \r| \le \del$ on the event $O_{\tau(\del)}$. Finally, we recall that \[\P[\O_{\tau(\del)}] \geq 1-4\exp\left(-\frac{\del^2 \log^2 L}{2\pi} \cdot \l(1-\frac{14K}{\del \sqrt{\log L}}\r)^2 \right). \]
	This completes the proof of the theorem.
\end{proof}

\section{Conclusion and future work}
In this paper, we have investigated signal analysis via an examination of the stochastic geometric properties of spectrogram level sets. We have obtained rigorous theorems demonstrating the efficacy of a spectrogram level sets based approach to the detection and estimation of signals, and further proposed a level sets based algorithm for signal analysis that is intrinsic to given spectrogram data. We hope that the present work, to our knowledge the first to establish rigorous statistical guarantees for spectrogram based methods for signal analysis, will serve as a prototype and provide a roadmap for similar investigations with regard to broader classes of signals. While the ambit of our results is substantiated by the fact that a general signal can in principle be written as a linear combination of Hermite functions, other bases might allow for more parsimonious signal representations, and thus be of practical interest as well. This includes the setting of complex-valued (analytic) chirps, which is a natural avenue for future investigation.

\section*{Acknowledgment}
We are grateful to R\'emi Bardenet for illuminating discussions. We thank the anonymous referees for their careful reading of the paper, and for their helpful comments and suggestions.

\end{document}